\documentclass[%
twocolumn,
superscriptaddress,
 amsmath,amssymb,
 aps,
 pra,
floatfix,
]{revtex4-2}

\usepackage{graphicx}
\usepackage{dcolumn}
\usepackage{bm}
\usepackage[bookmarks=false,colorlinks,citecolor=red]{hyperref}
\usepackage{color}
\usepackage{dcolumn} 
\usepackage{bm}      
\usepackage{array}   
\usepackage{amssymb}
\usepackage{amsmath}
\usepackage{amsfonts}
\usepackage{hyperref}
\usepackage{url}
\usepackage[T1]{fontenc}
\usepackage{multirow}
\usepackage{booktabs}
\usepackage[british]{babel}
\usepackage{times}
\usepackage{chemformula}
\usepackage{tikz} 
\usepackage[colorlinks,citecolor=red]{hyperref}
\usepackage{qcircuit}
\usepackage{mathtools}
\usepackage{subcaption}
\usepackage{enumerate}
\usepackage{lipsum}
\usepackage{caption}

\DeclareMathOperator{\atantwo}{atan2}

\newcommand{\ket}[1]{\left | #1 \right \rangle}

\include{MyCommand}

\begin{document}

\title{Constructing Qudits from Infinite Dimensional Oscillators by Coupling to Qubits} 

\author{Yuan Liu}
 \affiliation{Department of Physics, Co-Design Center for Quantum Advantage, Massachusetts Institute of Technology, Cambridge, Massachusetts 02139, USA} 

\author{Jasmine Sinanan-Singh}
 \affiliation{Department of Physics, Co-Design Center for Quantum Advantage, Massachusetts Institute of Technology, Cambridge, Massachusetts 02139, USA}
 
\author{Matthew T. Kearney}
 \affiliation{Department of Electrical Engineering and Computer Science, Massachusetts Institute of Technology, Cambridge, Massachusetts 02139, USA}
 
\author{Gabriel Mintzer}
 \affiliation{Department of Physics, Massachusetts Institute of Technology, Cambridge, Massachusetts 02139, USA}
 \affiliation{Department of Electrical Engineering and Computer Science, Massachusetts Institute of Technology, Cambridge, Massachusetts 02139, USA}
 
\author{Isaac L. Chuang}
 \affiliation{Department of Physics, Co-Design Center for Quantum Advantage, Massachusetts Institute of Technology, Cambridge, Massachusetts 02139, USA}
 \affiliation{Department of Electrical Engineering and Computer Science, Massachusetts Institute of Technology, Cambridge, Massachusetts 02139, USA}
 
\date{\today}

\begin{abstract}
An infinite dimensional system such as a quantum harmonic oscillator offers a potentially unbounded Hilbert space for computation, but accessing and manipulating the entire state space requires a physically unrealistic amount of energy.  When such a quantum harmonic oscillator is coupled to a qubit, for example via a Jaynes-Cummings interaction, it is well known that the total Hilbert space can be separated into independently accessible subspaces of constant energy, but the number of subspaces is still infinite.  Nevertheless, a closed four-dimensional Hilbert space can be analytically constructed from the lowest energy states of the qubit-oscillator system.  We extend this idea and show how a $d$-dimensional Hilbert space can be analytically constructed, which is closed under a finite set of unitary operations resulting solely from manipulating standard Jaynes-Cummings Hamiltonian terms. Moreover, we prove that the first-order sideband pulses and carrier pulses comprise a universal set for quantum operations on the qubit-oscillator qudit. This work suggests that the combination of a qubit and a bosonic system may serve as hardware-efficient quantum resources for quantum information processing.
\end{abstract}

\maketitle


\section{\label{sec:intro} Introduction}

    Quantum harmonic oscillators are promising resources for quantum computation, owing to the infinite number of available states and their ubiquitous presence in nature as molecular vibrations \cite{tesch2002quantum}, solid-state phonons \cite{ashcroft1976solid}, and optical/microwave cavities \cite{blais2020quantum}. The high dimensionality of oscillators provides not only a starting point for various bosonic quantum error-correcting codes 
    \cite{chuang1997bosonic,gottesman2001encoding,cochrane1999macroscopically,michael2016new,niu2018hardware,albert2018performance,noh2020encoding},
    but also a natural physical platform for universal quantum computation \cite{gottesman1999fault,zhou2003quantum,niu2018qudit}. It is therefore desirable to achieve universal control over quantum harmonic oscillators. However, direct transitions driven between Fock states of a single oscillator will leak states outside of any finite computational space of an oscillator, due to its equally spaced and open ended spectrum. This means controlling the entire infinite dimensional Hilbert space of an oscillator requires an infinite amount of energy and is thus highly unphysical.

    In contrast to the infinite dimensions of oscillators, quantum computation often relies on a closed \emph{finite} dimensional Hilbert space to represent quantum states and perform unitary operations on these states such that they remain inside the computational space \cite{nielsen2010quantum}. By truncating the oscillator to a low energy subspace and performing computation inside this finite dimensional subspace, the unphysical requirement to the amount of control resources can be alleviated. It is both satisfying and somewhat surprising that the most elegant way to truncate an oscillator is to couple it to a qubit (a two-level system), the simplest primitive of quantum computation. 
    The conversions between continuous-variable \cite{braunstein2005quantum} and discrete-variable states enabled by the qubit+oscillator systems have spurred many important developments in the field. On the one hand, such qubit+oscillator systems have been used to realize various quantum error-correcting codes by encoding finite dimensional qubits into continuous-variable bosonic modes, including the GKP code \cite{pirandola2006continuous,motes2017encoding,fluhmann2019encoding}, the binomial code \cite{michael2016new}, and the cat code \cite{ofek2016extending}. On the other hand, it has been shown that continuous-variable states can be transferred into multi-qubit states using primitive operations common to trapped ion systems \cite{hastrup2021universal}.
    
    There are different ways to couple a qubit to an oscillator, and one of the most common couplings is described by the Jaynes-Cummings (JC) interaction \cite{shore1993jaynes} due to its broad applicability.
    Based on the JC Hamiltonian, Law and Eberly \cite{law1996arbitrary} demonstrated how arbitrary single-mode Fock states can be prepared by using sideband and carrier transitions in an alternating fashion. This idea was further extended by Mischuck and Mølmer to synthesize arbitrary unitary operations in a qubit-oscillator system by decomposing the unitary into many state-preparation protocols \cite{mischuck2013qudit}. A different approach was taken to achieve universality for a resonantly coupled superconducting cavity to an artificial atom, despite the requirement of a slow adiabatic crossover for cavity states from the coupled to the uncoupled regime \cite{strauch2012all}. Moreover, universal control over an oscillator is also discussed beyond the JC interaction. For example, a simple quantum circuit was proposed to realize universal control by carefully engineering a partially resonant and partially dispersive coupling between an auxiliary three-level system and an oscillator \cite{santos2005universal}. Universal control was also demonstrated for an oscillator coupled to a qubit fully dispersively \cite{krastanov2015universal}, by combining a selective number-dependent arbitrary phase operation with a displacement operation on the oscillator.
    
    In the various protocols developed over the past two decades on universal control over oscillators, the question of how to close an arbitrary finite dimensional subspace of an oscillator was investigated far less frequently. Childs and Chuang \cite{childs2000universal} showed that the lowest two levels of the oscillator can be closed to form a four dimensional Hilbert space where arbitrary unitary operations can be realized. This closed, truncated space has been used to experimentally implement the Deutsch-Josza algorithm on a trapped ion quantum computer \cite{gulde2003implementation,schmidt2003realize}. One of the key ideas in their construction is to synchronize the rotations on the two Bloch spheres associated with the two 2-dimensional subspaces, using dynamical decoupling by a four-pulse sequence adapted from nuclear magnetic resonance \cite{vandersypen2005nmr}.
    
    The more recent work of Mischuck and Mølmer \cite{mischuck2013qudit} constructed arbitrary unitary operations in the lowest $(n+1)$-dimensional subspace of an oscillator (for any $n \in \mathbb{Z}$) 
    by decomposing the unitary into a series of modified state preparation protocols. Each state preparation unitary is then synthesized without leaking states in the computational space to the outside. They further proved arbitrary two-qudit gates are possible by coupling two oscillators to the same qubit, enabling qudit-based quantum computation. By synthesizing arbitrary independent sideband transitions in each sideband subspace using a truncated Fourier series on the rotation angles \cite{pryor2006fourier}, they demonstrated the ability to realize an arbitrary unitary on the $(n+1)$-dimensional subspace, using ~$O(n^{18.5} \delta^{-3})$ pulses (together with a large prefactor), where $\delta$ is the error of the synthesized unitary with respect to the target unitary. This employs a powerful technique from optimal control theory, with further efficiencies gained through numerical optimization. On the other hand, the use of optimal control theory and optimization renders the protocol \emph{approximate} in nature. 
    Similar ideas for truncating the oscillators has been employed in the context of perfectly generating atomic coherence from optical coherence in a recent work \cite{goldberg2020transcoherent}.
    
    In the present work, we address the same challenge raised by \cite{mischuck2013qudit}, but we seek to solve the problem of closing off the $(n+1)$-dimensional oscillator subspace using a formalism which is fully analytical, sans results from optimal control theory or optimization.  This allows us to fully understand the algebraic structures present in the JC Hamiltonian and how they may be exploited with explicit algorithms. Also, having protocols which are exact opens the door to understanding trade-offs and potential impact of errors. In principle, the powerful optimization techniques of \cite{mischuck2013qudit} could also be deployed on top of a fully analytical solution to improve its scaling. Moreover, from the viewpoint of control theory, we would like our control set to be as simple as possible. Ideally, we want to use a finite set of basic control operations (e.g. laser frequencies) regardless of the dimension of the computational space, as opposed to the constructions such as those employing a dispersive coupling Hamiltonian where the number of control frequencies increases linearly with the dimension of the computational space \cite{krastanov2015universal}.
    
    We produce an analytical solution by building on the approach of Ref. \cite{childs2000universal} to systematically close off an arbitrary $(n+1)$-dimensional low energy subspace of an infinite dimensional oscillator via coupling to a single qubit; further, we construct universal unitary operations within this subspace \emph{fully analytically} using only first-order sideband and carrier pulses. This analytical construction of a unitary relies on exploiting algebraic structures in the problem. The key structures arise from the well-known fact that the full Hilbert space of the qubit+oscillator is naturally partitioned into an infinite number of 2-dimensional subspaces, and only the subspace at the boundary leaks states in the computational space to the outside. We exploit these structures by constructing a set of elementary SU(2) rotations on these arrays of 2-dimensional subspaces as the instruction set for constructing arbitrary unitaries. We give a recursive protocol to construct such elementary SU(2) rotations by cleaning each 2D subspace one by one without accumulating any errors. Due to this recursive fashion, we refer to the construction process as \emph{recursive cleaning} and designate the qudits constructed as \emph{qubit-oscillator qudits} (QO-qudits). It should be noted that our recursive cleaning construction is applicable to a larger class of Hamiltonian beyond the JC interaction, but we will use the JC interaction for the ease of discussion. Also, note that the qubit doubles the Hilbert space of the truncated oscillator resulting in a larger $d = 2(n+1)$ dimensional space. We prove universal control on this enlarged space, and this naturally implies universal control on the $(n+1)$-dimensional truncated oscillator, which is a subspace of the enlarged $d$-dimensional space. We shall use the notation $d$-QO-qudit to represent a qubit coupled to a truncated oscillator with Fock levels $\{ \ket{0}, \ket{1}, \ket{2}, ..., \ket{n} \}$. This convention is consistent with previous work in Ref. \cite{childs2000universal}, despite doubling the dimension of the computational space in Refs. \cite{krastanov2015universal,mischuck2013qudit}.
    
    Our results are presented as follows. In Sec. \ref{sec:model}, we introduce the qubit-oscillator coupling Hamiltonian and derive the available unitary operations, which serves as a starting point for the rest of the discussions. We then describe a general strategy for finding unitary operations that are closed for arbitrary $d$-QO-qudits in Sec. \ref{sec:close-off}. A constructive universality proof of the QO-qudit operations is given in Sec. \ref{sec:universality}. We give an explicit algorithm for our construction and show an example for an 8-QO-qudit in Sec. \ref{sec:examples}, where the theoretical bound on the number of pulses are also supported numerically up to $46$-QO-qudit. Finally, we summarize in Sec. \ref{sec:conclusion} and discuss potential future directions to explore.

\section{Theoretical Model for the QO-Qudits \label{sec:model}}

For completeness, we give a short review of the Hamiltonian used in our analysis. The total Hamiltonian of the system is given by $H = H_0 + H_I$, where $H_0 = \hbar \omega_{0} \sigma_z / 2 + \hbar \omega_z a^\dagger a $ is the non-interacting Hamiltonian of the qubit and oscillator. The interaction between the qubit and the oscillator can be described by a spin-{1/2} particle interacting with an electromagnetic (EM) field,
\begin{align}
    H_I = - \vec{\mu} \cdot \vec{B} ,
    \label{int-Hamiltonian}
\end{align}
where $\vec{\mu} = \mu \vec{\sigma}/2$ is the magnetic moment and $\vec{B} = B \hat{x} {\rm cos}(kz - \omega t + \Phi)$ is the magnetic field associated with the external drive. In the second quantized form, the position $z = z_0 (a + a^\dagger)$, where $z_0$ is the characteristic length of the oscillator's ground state motional wave function, $a$ and $a^\dagger$ are the annihilation and creation operators of the oscillator.

Under the dipole approximation $ \eta \equiv kz_0 \ll 1$ and abandoning the fast rotating terms, we can expand Eq. \eqref{int-Hamiltonian} into power series of $\eta$. Depending on the frequency of the external drive $\omega$, the following unitary operations that couple different states of the QO-qudits may be implemented. Denote the computational basis of the QO-qudit as $\{ \ket{\alpha, n} \}$ where $\alpha = 0, 1$ labels the two qubit states, and $n = 0, 1, 2, ...$ labels the oscillator levels. When $\omega = \omega_0$, the EM field couples the lower and upper levels of the qubit directly, leading to a carrier pulse $V_c$ that performs the same rotation $P(\theta, \phi)$ in each subspace $\{ \ket{0, n}, \ket{1, n} \}$
\begin{align}
    V_c (\theta, \phi) = {\rm exp}\Big[ i \frac{\theta}{2} (e^{i\phi} \sigma^+ + e^{-i\phi} \sigma^-) \Big] = \bigoplus_{n=0}^{\infty} P(\theta,\phi),
    \label{carrier-transition}
\end{align}
where $P(\theta,\phi)$ is a single qubit rotation around an axis with angle $\phi$ in the $xy$-plane by an amount $\theta$. 
We can likewise couple each pair of $\{ \ket{0, n}, \ket{1, n-1} \}$ ($n=1,2,...$) states using the first-order red sideband transition by setting $\omega = \omega_0 -  \omega_z$, leading to
\begin{align}
    V_s (\theta, \phi) = {\rm exp}\Big[ i \frac{\theta}{2} (e^{i\phi} \sigma^+ a + e^{-i\phi} \sigma^- a^{\dagger}) \Big] = \bigoplus_{n=1}^\infty Q_n(\theta,\phi).
    \label{sideband-transition}
\end{align}
In the above, $\sigma^{\pm} = (\sigma_x \pm i \sigma_y)/2$ and $\sigma_x, \sigma_y, \sigma_z$ are the Pauli operators of the qubit. The rotation angle $\theta$ and rotation axis $\phi$ are given by
\begin{align}
    \theta &= - \frac{\mu B t \eta^m}{2 \hbar m!}, \\
    \phi &= \Phi + (m ~~{\rm mod}~ 4) \frac{\pi}{2}, \label{pulse-phase}
\end{align}
with $m=0$ for the carrier pulse and $m=1$ for the first-order sideband pulse, where $t$ is the pulse duration. Note the red sideband performs a $n$-dependent rotation $Q_n$ along the same axis defined by $\phi$ but by a different rotation angle on each 2D subspace. The first-order blue sideband as well as higher order sideband transitions can be similarly derived. Since the blue sideband pulse can be obtained by conjugating the red sideband pulse using a carrier $\pi$-pulse, we only need to consider the carrier and the red sideband pulses. In the following, we will show that $V_c$ and $V_s$ as defined in Eqs. \eqref{carrier-transition} and \eqref{sideband-transition} are sufficient to generate arbitrary unitary operations in QO-qudits. 

\begin{figure}[ht]
    \begin{center}
        \includegraphics[width=\columnwidth]{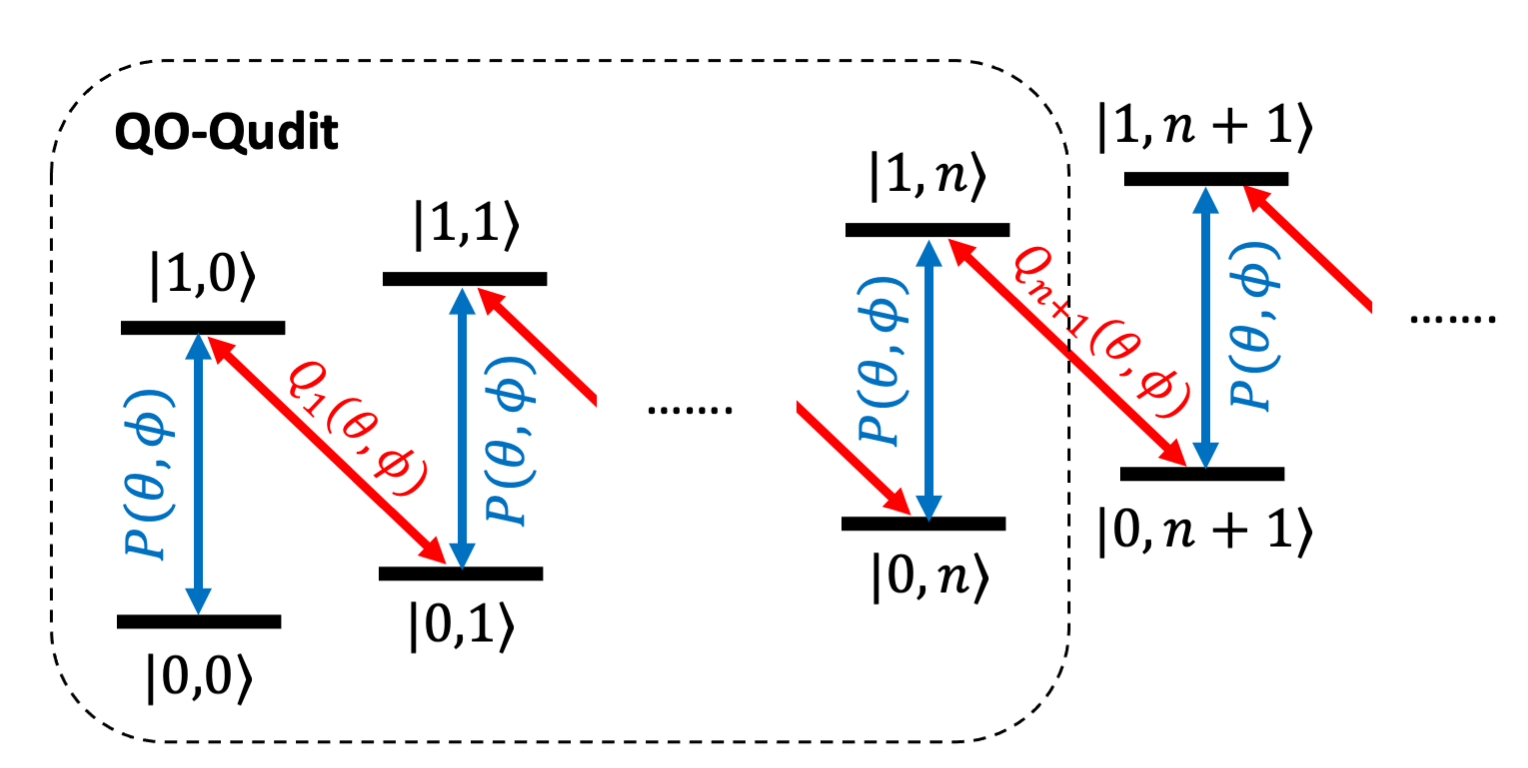}
    \end{center}
    \caption{Energy level diagram in a $d$-QO-qudit. The first index in the state label represents the qubit state and the second represents the oscillator Fock level. Red arrows indicate states that are coupled through red sideband transitions (forming the sQM), while blue arrows indicate states coupled through the carrier pulse (forming the cQM).}
    \label{fig:qoqudit}
\end{figure}

An energy level diagram of the qubit-oscillator system as well as relevant transitions are labeled in Fig. \ref{fig:qoqudit}, with the definition of a $d$-QO-qudit explicitly shown. To simplify the discussion, we partition the Hilbert space of a $d$-QO-qudit into sets of two-dimensional subspaces in two different ways based on the action of the red sideband pulse and the carrier pulse. Each two-dimensional subspace can be viewed as a single qubit. The sideband qubit manifold (sQM) are spanned by states $\{\ket{0,j}, \ket{1,j-1}\}_{j=1}^{n}$ where each subspace is labeled by its corresponding value of $j$. We do not include states $\ket{0,0}$ and $\ket{1, n}$ in sQM because they remain intact during the sideband operation (up to a parity phase). The carrier qubit manifold (cQM) are spanned by $\{\ket{0,j}, \ket{1,j}\}_{j=0}^{n}$, where $j$ start from 0 instead of 1. According to the above definition, there are $n$ nontrivial 2D subspaces in the sQM and $(n+1)$ nontrivial 2D subspaces in the cQM. To differentiate the unitary operations on these two qubit manifolds, we will use a tilde when referring to the cQM.

\section{Close Off an Infinite Dimensional Harmonic Oscillator \label{sec:close-off}}

The first step toward constructing a QO-qudit is to close off a low energy subspace of the oscillator, which serves as the computational space. The closedness of a computational space means that any the states inside it will not leak to any states outside while performing any quantum computations. This is an important step to accomplish since any leakage will result in a non-unitary evolution of the computational space and ruin the computation immediately. In general, this is difficult to achieve because there are infinitely many states outside of the finite computational space for an oscillator, which suggests that we may need to eliminate infinitely many coupling amplitudes between the computational space and its orthogonal space. This general condition, nevertheless, is greatly simplified in our case due to the partition of the full Hilbert space into many 2D subspaces. Moreover, since each unitary operation may leak states outside, there is leakage possibly at every step in performing quantum computation as is the case in Ref. \cite{mischuck2013qudit}. We take a different approach here by selecting a subset of unitary operations such that the closedness of the computational space is always guaranteed. We will describe in detail how this can be accomplished in the following.

\subsection{Close Off the Transition on the Boundary}
From Fig. \ref{fig:qoqudit}, we first note that all carrier transitions $P(\theta,\phi)$ do not leak states in the $d$-QO-qudit outside for any $d$. As a result, we need only focus on the sideband pulses. The only sideband transition that leaks states is the $Q_{n+1}(\theta,\phi)$ transition on the boundary that couples the $\ket{1, n}$ state inside to the state $\ket{0, n+1}$ outside of the QO-qudit. It is therefore sufficient to shut off the coupling between these two states. We can calculate explicitly the matrix elements of $Q_{n+1}$ to be
\begin{align}
    Q_{n+1}(\theta, \phi) = 
    \begin{bmatrix}
        {\rm cos} \theta_{n+1} & i e^{-i\phi} {\rm sin} \theta_{n+1} \\
        i e^{i\phi} {\rm sin} \theta_{n+1} & {\rm cos} \theta_{n+1}
    \end{bmatrix},
\end{align}
where $\theta_{n+1} = \frac{\sqrt{n+1}}{2} \theta$. When driving a multiple of $2\pi$ pulse from $\ket{0, n+1}$ to $\ket{1, n}$, i.e., $\theta_{n+1} = \pi k$ where $k$ is an integer, we can decouple these two states since ${\rm sin}\theta_{n+1} = 0$.  The smallest $\theta$ (shortest pulse, $k=1$) that can realize this decoupling is
\begin{align}
    \theta = \frac{2\pi}{\sqrt{n+1}},
    \label{theta_decouple}
\end{align}
which gives
\begin{align}
    Q_{n+1}(\frac{2\pi}{\sqrt{n+1}}, \phi) = 
    \begin{bmatrix}
        -1 & 0 \\
        0 & -1
    \end{bmatrix}.
\end{align}
This constraint also guarantees that the operation on $\ket{1,n}$ is trivial (up to a parity phase), which is why we exclude it from the subspaces of the sQM.

\subsection{Conjugacy Class}
It may be argued that the above requirement on $\theta$ to be discrete special values seems to significantly constrain the number of possible unitaries we can apply, and thus hampers the universality of the QO-qudit. However, this is not the case because conjugating a closed sideband pulse using an arbitrary unitary operation leads to another closed unitary operation, i.e., an operation $F = U^\dagger G U$ will be closed for an arbitrary unitary $U$ if the given unitary $G$ is closed. In the case of SU(2), $F$ will be a rotation on the Bloch sphere by the same angle $\theta$ as the original rotation $G$, but along a different axis defined by $U$. This concept of the conjugacy class of a given unitary operation provides us enough flexibility to construct new unitaries and is the key idea behind our recursive cleaning proof, which we shall discuss in detail in the next section.

\section{Universality Proof \label{sec:universality}}


With a closed computational space (QO-qudit) established, we next construct arbitrary unitary operations on this subspace using the operations given in Eqs. \eqref{carrier-transition} and \eqref{sideband-transition} subject to the closedness condition, which we refer to as the universality of the QO-qudit. We prove the QO-qudit universality in this section in two different pictures. We give some intuition of the proof in the Hamiltonian picture in Sec. \ref{sec:ggms-universal}. We then move to the unitary picture and give explicit constructions for arbitrary QO-qudit unitary operations in two steps, as is described in Sec. \ref{sec:cleanElemSU2ToUniversality} and \ref{sec:cleanElemSU2Construct}. First, we reduce the QO-qudit universality to \emph{elementary} SU(2) rotations in sQM in Sec. \ref{sec:cleanElemSU2ToUniversality}. We then give explicitly constructions for such elementary SU(2) rotations in the sQM in Sec. \ref{sec:cleanElemSU2Construct}.

\subsection{Intuition in the Hamiltonian Picture \label{sec:ggms-universal}}

One of the well-known criteria for the universality of qudit-based quantum computation is the ability to perform arbitrary SU(2) rotations between any two levels in the qudit space \cite{gottesman1999fault,brennen2005criteria}. 
In the Hamiltonian picture, this means we need to obtain the full $\mathfrak{su}(d)$ Lie algebra with $(d^2-1)$ elements that generates the $SU(d)$ group of the $d$-QO-qudit. Matrix representations of such generators are also known as the generalized Gell-Mann matrices (GGMs) \cite{bertlmann2008bloch} which are all Hermitian (See Appendix \ref{app:definitionGM} for a definition of GGMs). In the special case of a qubit, $d=2$, there are three generators, i.e., the Pauli matrices. Similar to the qubit case, these GGMs can be classified into three categories denoting the rotation of $x, y,$ and $z$ type, respectively. Moreover, the $z$ type GGMs may be obtained by multiplying the $x$ and $y$ types. We will give an argument on how to generate all the GGMs of a single $d$-QO-qudit. The aim of this section is mainly to convey an intuition of our proof. In the next several sections we will provide a rigorous proof in the unitary picture. 

As we noted before, the first observation from Fig. \ref{fig:qoqudit} is that the sideband pulse (or the carrier pulse) naturally partitions the $d$-QO-qudit into $n$ small 2D subspaces (again states $\ket{0,0}$ and $\ket{1,n}$ are discarded). In other words, the Hamiltonian that generates the sideband pulse can be written as a direct sum of $n$ GGMs. By controlling the phase in Eq. \eqref{pulse-phase}, the $n$ GGMs can either be of $x$ or $y$ type. Therefore, one simple idea to generate a single clean GGM is to cancel the other $(n-1)$ GGMs in the direct sum by dynamical decoupling. The $\sqrt{n}$ dependency of the Rabi frequency for each subspace in the sideband pulse provides a possibility for such decoupling to be done. To see this more clearly, imagine that we have the GGM $H_n$ for the $d$-QO-qudit in the sideband manifold
\begin{align}
    H_n &= 0_1 \oplus ... \oplus \Sigma_k \oplus ... \oplus 0_{n} \oplus h_{n+1} ...,
    \label{h_n}
\end{align}
where 
$0_n$ is a null Hamiltonian in the $n$-th subspace,
$\Sigma_k = \{ \sigma_x, \sigma_y \}$, $h_{n+1} = \bm{\hat{r}}_{n+1} \cdot \bm{\sigma}$ is some arbitrary Hamiltonian in the $(n+1)$-th subspace defined by a unit vector $\bm{\hat{r}}_{n+1}$. Our goal is then to create the corresponding GGM $H_{n+1}$ for the $(d+2)$-QO-qudit
\begin{align}
    H_{n+1} &= 0_1 \oplus ... \oplus \Sigma_k \oplus ... \oplus 0_{n} \oplus 0_{n+1} \oplus h_{n+2} ...
    \label{h_n+1}
\end{align}
Recall that we can always flip the sign of a single-qubit Hamiltonian by conjugating it using an $SU(2)$ rotation along an axis that is perpendicular to the Hamiltonian. One simple example of such phase flip is $X^\dagger Z X = -Z$, where $X = i\sigma_x$. We may therefore use such dynamical decoupling trick to flip the phase of $h_{n+1}$ in Eq. \eqref{h_n} by conjugating it with a set of red sideband pulse $V_s$ such that
\begin{align}
    h_{n+1} + (V_s^\dagger)_{n+1} h_{n+1} (V_s)_{n+1} = h_{n+1} - h_{n+1} = 0.
\end{align}
This reproduces the $0_{n+1}$ term in Eq. \eqref{h_n+1}. This conjugation of course also alters $\Sigma_k$ in Eq. \eqref{h_n} into a different Hamiltonian, but the deviation from $\Sigma_k$ may be cleaned again by conjugating using another set of red sideband pulses to fully recover Eq. \eqref{h_n+1}. These step can be repeated recursively to obtain GGMs for any $n$ in the sQM manifold, and this is also the main technique we will use in our proof in the next few sections.





The decoupling described above will enable us to generate GGMs that couple two adjacent states $\{ \ket{0, j}, \ket{1, j-1} \}$, but it remains to generate other GGMs that couple states far from each other. This leads to our second observation from Fig. \ref{fig:qoqudit}: alternatively applying the GGMs in the sQM and those in the cQM can couple states far apart together, provided that we can access the cQM GGMs. We will show in the next section how rotations in the cQM can be generated from those in the sQM. These GGMs then give us the universality for the full $SU(d)$ group of a $d$-QO-qudit.



\subsection{From Clean Elementary SU(2) Rotations in sQM to QO-Qudit Universality \label{sec:cleanElemSU2ToUniversality} }

In this section, we will start from the qudit criteria in Ref. \cite{gottesman1999fault,brennen2005criteria}, and show that arbitrary SU(2) rotations between any two levels in the QO-qudit can be obtained from a set of clean elementary SU(2) rotations in sQM alone. This is accomplished in two steps as follows. First, in Sec. \ref{sec:cleanArbSU2ToUniversality}, we reduce the QO-qudit universality to the ability to construct \emph{arbitrary clean} SU(2) rotations in \emph{both} sQM and cQM. Secondly, by a basis transformation from the sQM to the cQM, combining with dynamical decoupling \cite{vandersypen2005nmr}, we show that constructing \emph{arbitrary clean} SU(2) rotations in both sQM and cQM can be further reduced to the construction of \emph{elementary clean} SU(2) rotations \emph{solely} in the sQM in Sec. \ref{sec:sQMTocQM}. 

\subsubsection{Arbitrary clean SU(2) rotations in sQM and cQM implies QO-qudit universality} \label{sec:cleanArbSU2ToUniversality}

We define an arbitrary clean SU(2) rotation in the $k$-th subspace of sQM for a $d$-QO-qudit ($d = 2(n+1)$) as
\begin{align}
    U_n^{(k)} = I_1 \oplus I_2 \oplus ... \oplus W_k \oplus ... \oplus I_n ,
    \label{sQM-clean-arb}
\end{align}
which performs a nontrivial arbitrary SU(2) rotation $W_k$ in the $k$-th sideband subspace and leaves the other sideband qubits unchanged, i.e., acted by an identity operation. Similarly, an arbitrary clean SU(2) rotation in the $k$-th subspace of the cQM is likewise denoted as
\begin{align}
    \tilde{U}_n^{(k)} = \tilde{I}_0 \oplus \tilde{I}_1 \oplus ... \oplus \tilde{W}_k \oplus ... \oplus \tilde{I}_n ,
    \label{cQM-clean-arb}
\end{align}
where tilde is used to distinguish it from the case of sQM above. Also, note that $\tilde{U}_n^{(k)}$ is a direct sum of $(n+1)$ carrier qubits, while $U_n^{(k)}$ is composed of $n$ sideband qubits, which is evident from Fig. \ref{fig:qoqudit}. 

For the QO-qudit universality, we require arbitrary unitary operations $V_t$ between any two levels, say $\ket{\alpha,p}$ and $\ket{\beta,q}$, where $p \le q$ without loss of generality and $\alpha, \beta = 0 \text{ or } 1$. In the following, we give explicit constructions in the case of $\alpha = 0, \beta = 1$; the other three cases ($\alpha = \beta = 0, \alpha = \beta = 1, \alpha = 1 \text{ and } \beta = 0$) may be constructed in a similar way. 

If $p=q$, $V_t$ simply corresponds to a clean SU(2) rotation $\tilde{U}_n^{(p)}$ in the cQM given by Eq. \eqref{cQM-clean-arb}. In the case of $p < q$, there is no single rotation in Eq. \eqref{sQM-clean-arb} or \eqref{cQM-clean-arb} that can produce the coupling between $\ket{0,p}$ and $\ket{1,q}$. However, we may first realize a unitary $\tilde{U}_n^{(p)}$ such that the SU(2) rotation in its $p$-th subspace $\tilde{W}_p$ satisfies $\tilde{W}_p = V_t$. We may then perform a sequence of Pauli X gates to swap the state $\ket{1,p}$ with $\ket{0,p+1}$, and then $\ket{0,p+1}$ with $\ket{1,p+1}$ and so on, until finally $\ket{1,q}$ is reached. Such a sequence of swap operations can be easily realized by chaining multiple clean SU(2) rotations in Eqs. \eqref{sQM-clean-arb} and \eqref{cQM-clean-arb} together. The overall pulse sequence to realize $V_t$ between $\ket{0,p}$ and $\ket{1,q}$ (with the rest state unaltered) is then
\begin{widetext}
\begin{align}
    [ U_n^{(p+1)} \tilde{U}_n^{(p+2)} ... U_n^{(q-1)} \tilde{U}_n^{(q)} ]^\dagger \cdot \tilde{U}_n^{(p)} \cdot [ U_n^{(p+1)} \tilde{U}_n^{(p+2)} ... U_n^{(q-1)} \tilde{U}_n^{(q)} ], 
\end{align}
\end{widetext}
where $W_k = \tilde{W}_k = X$ in all $U_n^{(k)}$ and $\tilde{U}_n^{(k)}$ for $k = p+1, p+2, ..., q-1, q$. $X = i \sigma_x$ is the usual Pauli X gate.

\subsubsection{Elementary clean SU(2) rotations in sQM implies arbitrary clean SU(2) rotations in sQM and cQM \label{sec:sQMTocQM}} 

In the above, we have shown that QO-qudit universality can be constructed from a set of clean arbitrary SU(2) rotations in sQM and cQM given by Eqs. \eqref{sQM-clean-arb} and \eqref{cQM-clean-arb}. Now we will show that such clean arbitrary SU(2) rotations may be constructed solely from a set of elementary clean SU(2) rotations in sQM alone.

We first define an \emph{elementary} clean SU(2) rotation $V_n^{(k)}$ in sQM (for the $k$-th subspace) for a $d$-QO-qudit ($d=2(n+1)$)
\begin{align}
    V_n^{(k)} = I_1 \oplus I_2 \oplus ... \oplus \Sigma_k \oplus ... \oplus I_n ,
    \label{sQM-clean-ele}
\end{align}
where $\Sigma_k = \{X, Y, I, -I\}$. It is now straightforward to see how an \emph{arbitrary} clean SU(2) rotation $U_n^{(k)}$ in Eq. \eqref{sQM-clean-arb} may be obtained from Eq. \eqref{sQM-clean-ele} and sideband transitions using refocusing. 

The basic idea of refocusing has been presented in the Hamiltonian picture in Sec. \ref{sec:ggms-universal}, we shall state it here again in the unitary picture here. For an arbitrary single qubit rotation $U(\theta, \phi)$ in the $xy$-plane, we first note that by conjugating with $Z$ gate, we can reverse the rotation direction $Z U(\theta, \phi) Z = U(-\theta,\phi)$. Therefore, the following pulse sequence would effectively cancel the effect of $U(\theta, \phi)$ and produce an identity operation
\begin{align}
    U(\theta,\phi) \cdot Z \cdot U(\theta,\phi) \cdot Z = I.
    \label{refocusing}
\end{align}
Now, note that an arbitrary rotation $W_k$ on the $k$-th subspace of sQM in Eq.  \eqref{sQM-clean-arb} can be decomposed as two rotations with rotation axis lying in the $xy$-plane (see Appendix \ref{app:decomp}). Therefore, it suffices to assume $W_n^{(k)}$ is a rotation with an axis in the $xy$-plane. Imagine we start from a sideband transition in Eq. \eqref{sideband-transition}, where $(\theta, \phi)$ are properly chosen such that the sideband rotation in the $k$-th subspace satisfies $Q_k(2\theta,\phi) = W_n^{(k)}$. And this also means for all other subspaces, there are nontrivial rotations that are not the identity operation. We will use refocusing to eliminate those unwanted rotations in the other subspaces as follows. From Eq. \eqref{sQM-clean-ele}, we may construct the following unitary $\bm{Z}_n^{(k)}$
\begin{align}
    \bm{Z}_n^{(k)} = Z_1 \oplus Z_2 \oplus ...\oplus Z_{k-1}  \oplus I_k \oplus Z_{k+1} \oplus ... \oplus Z_n ,
    \label{sQM-vecZ}
\end{align}
which applies a Pauli Z gate to each sideband qubit except the $k$-th one (trivially acted upon by identity). The following construction achieves Eq. \eqref{sQM-clean-arb}
\begin{align}
    U_n^{(k)} = V_s(\theta,\phi) \bm{Z}_n^{(k)\dagger} V_s(\theta,\phi) \bm{Z}_n^{(k)}, 
    \label{sQM-final-refocus}
\end{align}
since $Q_j(\theta,\phi) \cdot Z_j \cdot Q_j(\theta,\phi) \cdot Z_j = I$ for all $j \ne k$. While for the $k$-th subspace,
\begin{align}
    Q_k(\theta, \phi) \cdot I_k \cdot Q_k(\theta, \phi) \cdot I_k = Q_k(2\theta, \phi) = W_k.
\end{align}
This completes our construction of Eq. \eqref{sQM-clean-arb} from Eq. \eqref{sQM-clean-ele}.

\begin{figure}[ht]
    \centering
    \includegraphics[width=\columnwidth]{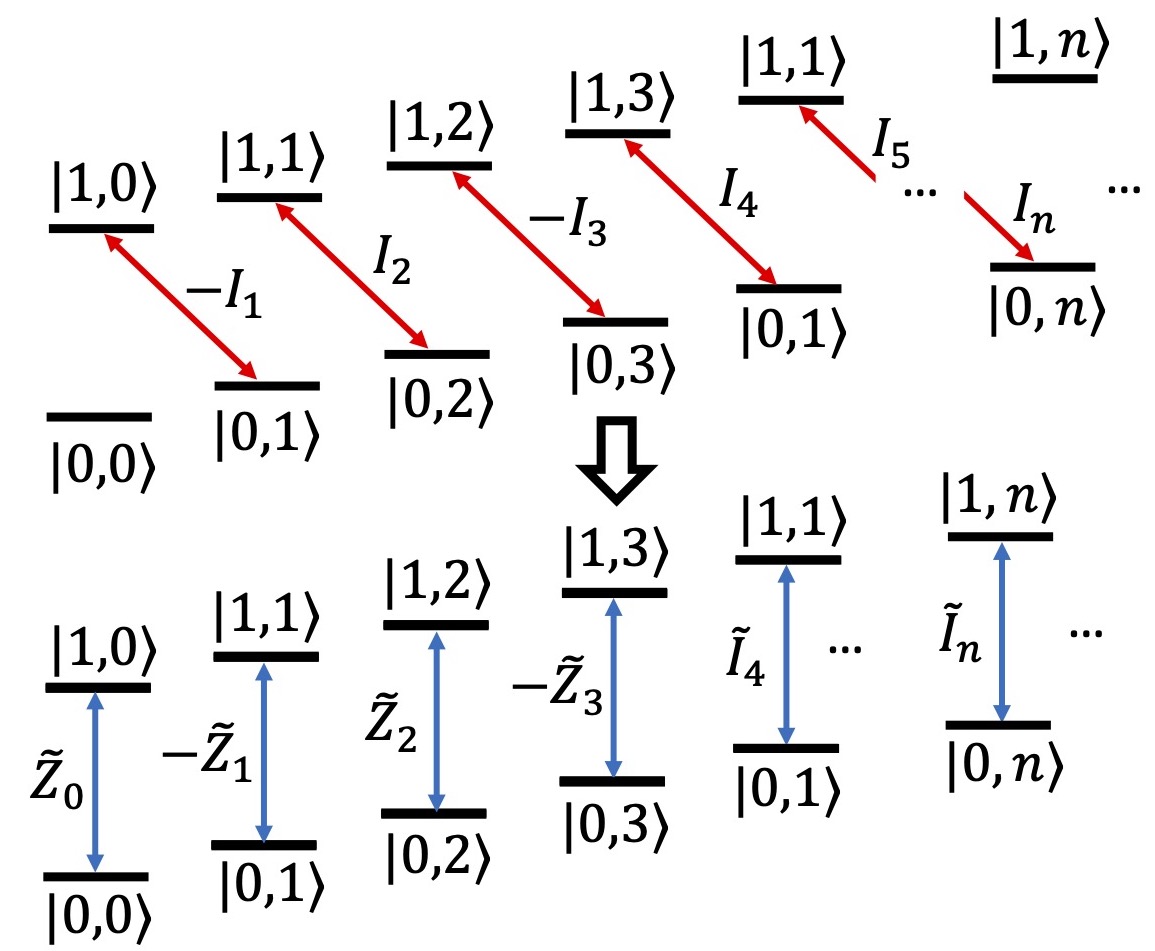}
    \caption{Translating between operations on sideband subspaces and carrier subspaces. The top energy diagram represents the unitary $-I_1 \oplus I_2 \oplus -I_3 \oplus I_4 ... \oplus I_n \oplus \dots$ acting on sideband subspaces which translates to the bottom energy diagram of $ \tilde{Z}_0 \oplus -\tilde{Z}_1 \oplus \tilde{Z}_2 \oplus -\tilde{Z}_3 \oplus \tilde{I}_4 ... \oplus \tilde{I}_n \oplus \dots$.}
    \label{fig:qm-translation}
\end{figure}

We now turn to construct arbitrary clean rotations in the cQM. This is facilitated by the fact that alternations of $\pm I$ in the sQM is equivalent to $I$ and $Z$ gates in the cQM up to a difference in the local parity. As an example, this is illustrated in Fig. \ref{fig:qm-translation} for the first four sideband qubits being $-I_1 \oplus I_2 \oplus -I_3 \oplus I_4$. They are equivalent to a gate sequence of $\tilde{Z}_0 \oplus -\tilde{Z}_1 \oplus \tilde{Z}_2 \oplus -\tilde{Z}_3$ for the first four carrier qubits.
In general, this conversion from the sQM to the cQM is described by

\begin{center}
 \begin{tabular}{||c| c c||} 
 \hline
$ \tilde{W}_{i-1}$ & $W_{i}$ & $W_{i-1}$ \\ [0.5ex] 
 \hline
\hline
 $\tilde{Z}$ & $-I$  & $+I$ \\ 
 -$\tilde{Z}$ & $+I$ & $-I$ \\
$\tilde{I}$ & $+I$ & $+I$ \\
 -$\tilde{I}$ & $-I$ & $-I$\\
 \hline
\end{tabular}
\end{center}

The above conversion also means that the $\{I, -I\}$ in each subspace from 1 to $(n+1)$ in sQM is equivalent to $\{\tilde{I}, \tilde{Z}\}$ in subspaces from 0 to $n$ in the case of cQM. Therefore, it is guaranteed that we can obtain the following unitary $\tilde{\bm{Z}}_n^{(k)}$ in cQM
\begin{align}
    \tilde{\bm{Z}}_n^{(k)} = \tilde{Z}_0 \oplus \tilde{Z}_1 \oplus ... \oplus \tilde{Z}_{k-1} \oplus \tilde{I}_k \oplus \tilde{Z}_{k+1} \oplus ...  \oplus \tilde{Z}_n ,
    \label{cQM-clean-ele}
\end{align}
which performs a Pauli Z gate on all carrier qubits of a $d$-QO-qudit, except the $k$-th qubit where it performs the identity. It follows that for a clean arbitrary unitary in Eq. \eqref{cQM-clean-arb} where the $k$-th subspace has a nontrivial rotation $\tilde{W}_k$ (assuming it is in $xy$-plane without loss of generality), we may choose a carrier pulse in Eq. \eqref{carrier-transition} with $(\theta, \phi)$ properly such that the carrier transition on each carrier qubit $P(2\theta,\phi) = \tilde{W}_k$. The following construction reproduces $\tilde{U}_n^{(k)}$
\begin{align}
    \tilde{U}_n^{(k)} = V_c(\theta,\phi) \tilde{\bm{Z}}_n^{(k)\dagger} V_c(\theta,\phi) \tilde{\bm{Z}}_n^{(k)}, 
    \label{cQM-final-refocus}
\end{align}
due to Eq. \eqref{refocusing} for all carrier subspaces except the $k$-th one, and for the $k$-th subspace,
\begin{align}
    P(\theta, \phi) \cdot \tilde{I}_k \cdot P(\theta, \phi) \cdot \tilde{I}_k = P(2\theta, \phi) = \tilde{W}_k.
\end{align}
This completes our construction of Eq. \eqref{cQM-clean-arb} from Eq. \eqref{sQM-clean-ele}. Specifically, by choosing $\tilde{W}_k$ to be a rotation on the $k$-th carrier qubit with a closed trajectory on its Bloch sphere, we can accumulate an arbitrary Berry phase on the $\ket{0,k}$ state. Combining many such Berry phase rotations together for different $k$, we immediately realize the SNAP gate in Ref. \cite{krastanov2015universal}.

\subsection{Construction for the Clean Elementary SU(2) Rotations in sQM \label{sec:cleanElemSU2Construct}} 

In this section, we give a \emph{recursive} proof on how to construct the \emph{elementary} clean SU(2) rotations in the sQM as in Eq. \eqref{sQM-clean-ele}. Our proof utilizes the repeated pattern of the oscillator's spectrum. In doing so, we first give the base case of a two-level oscillator coupled to a qubit, and then show how to clean up each subspace into the elementary operations for an $(n+1)$-level oscillator recursively. We also give bounds on the number of sideband pulses required in our construction.

\subsubsection{Recursive proof} \label{sec:inductiveProof}
Our claim is that we can construct arbitrary clean rotations upon any 2D subspace in the sQM manifold of a $d$-QO-qudit (again $d=2(n+1)$), $ V_n^{(k)} = I_1 \oplus I_2 \oplus \cdots \oplus \Sigma_k \oplus \cdots \oplus I_n \oplus \cdots$ with $\Sigma_k \in \{X, Y, -I\}$, $k \leq n$. When $n \leq 2$, we provide a direct construction.
For $n = 1$, we can easily select $(\theta, \phi)$ such that $V_1^{(1)} = V_s(\theta, \phi) = \Sigma_1$. For $n=2$, we will directly construct $I_1 \oplus X_2$ and $X_1 \oplus I_2$. Obtaining $Y$ instead of $X$ simply corresponds to mapping every red sideband pulse $V_s(\theta, \phi) \mapsto V_s(\theta, \phi + \frac{\pi}{2})$, and obtaining $-I$ instead of $X$ is done by repeating the sequence as $X^2 = -I$ (note we define $X = i\sigma_x$ throughout the paper). The following pulse sequences use conjugation to orchestrate clean rotations on the first two subspaces in 4 pulses: 
\begin{align}
    V_s\left(\sqrt{2}\pi, \phi_1 \right)V_s\left(\frac{\pi}{2}, 0\right) V_s\left(\sqrt{2}\pi, \phi_1 \right)V_s\left(-\frac{\pi}{2}, 0\right) \nonumber \\
    = X_1 \oplus I_2 \oplus ...\\
    V_s\left(2 \pi, \phi_2\right)V_s\left(\frac{\pi}{2\sqrt{2}},0\right) V_s\left(2 \pi, \phi_2\right)V_s\left(-\frac{\pi}{2\sqrt{2}},0\right) \nonumber \\ 
    = I_1 \oplus X_2 \oplus ...
    \label{four-pulse-IX}
\end{align}
where $\phi_1 = \cos^{-1}\left(\cot{\frac{\pi}{\sqrt{2}}}\right)$ and $\phi_2 = \cos^{-1}\left(\cot{\sqrt{2}\pi}\right)$. 

\begin{figure}
    \begin{center}
        \includegraphics[width=\columnwidth]{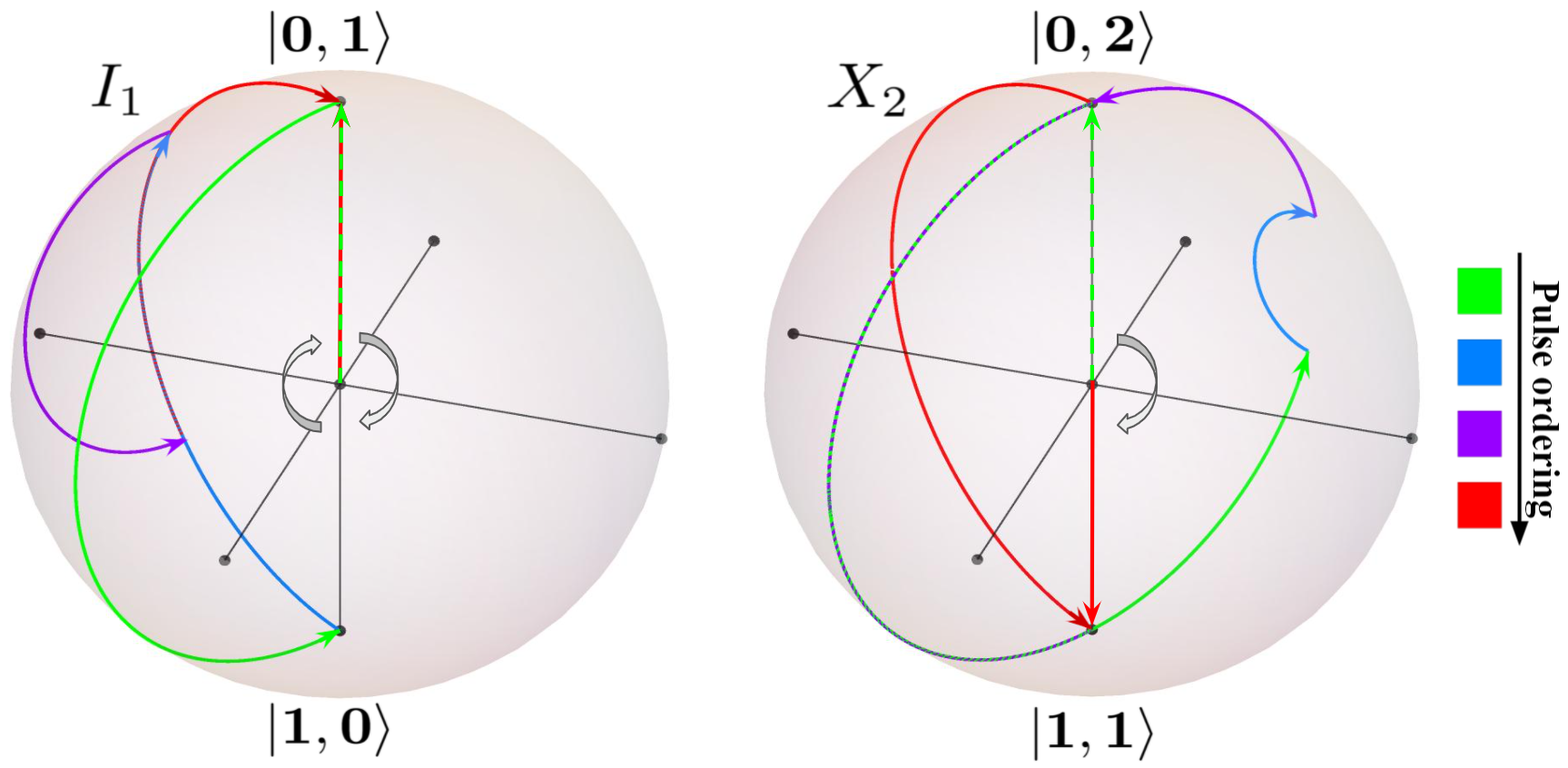}
    \end{center}
    \caption{
    A pulse sequence for $n = 2$ to realize $I_1 \oplus X_2$ in the first two sQM subspaces. Each Bloch sphere represents an sQM subspace spanned by the states at the North and South pole. The initial Bloch vector in dashed green (pointing towards North pole) is rotated to the final Bloch vector in solid red with the four pulse sequence from Eq. \eqref{four-pulse-IX}.}
    \label{fig:four-pulse-IX}
\end{figure}

An example pulse sequence for $n = 2$ is shown in Fig. \ref{fig:four-pulse-IX}. For $n > 2 $, we provide a recursive construction that also makes use of conjugation sequences to manipulate particular subspaces without disturbing others.

\textbf{Base Case} We begin by cleaning (taking to the identity) two subspaces $\mu_1, \mu_2 \in M = \{1, ..., n\} \setminus k$ with the constraint: 
\begin{align}
    \sqrt{\frac{k}{\mu_1}}, ~\sqrt{\frac{k}{\mu_2}} \notin  \mathbb{Z} .\label{mu12-constraint}
\end{align}
Using red sideband pulses $V_s(\theta, \phi)$, we can construct
\begin{widetext}
\begin{equation}
\begin{split}
    U_2 & = V_s\left(\frac{2 \pi}{\sqrt{\mu_1}}, 0\right) V_s\left(\frac{\pi}{\sqrt{\mu_2}}, \frac{\pi}{2}\right) V_s\left(\frac{2 \pi}{\sqrt{\mu_1}},0\right) V_s\left(\frac{-\pi}{\sqrt{\mu_2}}, \frac{\pi}{2}\right) \\
    &= \Omega_1^{(2)} \oplus \cdots \oplus I_{\mu_1} \oplus \cdots \oplus I_{\mu_2} \oplus \cdots \oplus \Omega_k^{(2)}\oplus \cdots \oplus \Omega_n^{(2)} \oplus \cdots
\end{split}
\label{u2}
\end{equation}
\end{widetext}
where the order of $\mu_1, \mu_2, k$ is arbitrary. Note that $U_2$'s subscript and $\Omega^{(2)}$'s superscript denote how many subspaces have been cleaned to the identity. With the constraint on $\mu_1, \mu_2$ in Eq. \eqref{mu12-constraint}, it is guaranteed that $\Omega_k^{(2)} \neq \pm I_k$. This is important since $\pm I$ each are the only element in their conjugacy class; if $\Omega_k = \pm I_k$, then we can never change $\Omega_k$ via conjugation. Our construction relies on using conjugated pulse sequences, so we must avoid $\Omega_k = \pm I_k$ at this step if we want it to be any other rotation. 

\textbf{Recursive step} Assume we have cleaned $j$ ($j<n$) subspaces indicated by the set $M_j = \{\mu_1, \mu_2, ..., \mu_{j}\}$. Then, ignoring order in the direct sum, we have  
\begin{equation}
    U_j = \bigoplus_{\mu_i \in M_j}I_{\mu_i}  \oplus \Omega_k^{(j)} \oplus \bigoplus_{m \in M \setminus M_j}\Omega_{m}^{(j)}
    \label{uj}
\end{equation}
We then choose a $\mu \in M \setminus M_j$ and clean $\Omega_\mu^{(j)}$ next. To do this we will move from $SU(2)$ into $SO(3)$ using the group homomorphism $R: SU(2) \xrightarrow{} SO(3)$ by identifying $\vec{s} \in \mathbb{R}^3$ with $s \in \mathfrak{su}(2)$ via $\vec{s} \leftrightarrow s = \vec{s} \cdot \vec{\sigma}$, with Pauli matrices $\vec{\sigma}$. Then, any $Q(\theta, \phi) \in SU(2)$ is mapped to a rotation about an axis $\vec{r}$ through angle $\theta$, $R_{\vec{r}}(\theta) \in SO(3)$ corresponding to $s \mapsto Q^\dagger s Q$.

In order to clean $R\left(\Omega_\mu^{(j)}\right) = R_{\vec{\mu}}(\theta_\mu)$, we choose an axis $\vec{\mu}_{\perp} \perp \vec{\mu}$ and note the following dynamical decoupling sequence
\begin{equation}
     R_{\vec{\mu}}(\theta_\mu) \left[ R_{\vec{\mu}_\perp}(\pi)  R_{\vec{\mu}}(\theta_\mu) R_{\vec{\mu}_\perp}(-\pi) \right] =  R_{\vec{\mu}}(\theta_\mu)  R_{\vec{\mu}}(-\theta_\mu)=  I,
\end{equation}
which cleans the $\mu$-th subspace. To find the red sideband pulses that perform $R_{\vec{\mu}_\perp}(\pi)$ on the $\mu$-th subspace, we must decompose $R_{\vec{\mu}_\perp}(\pi)$ into rotations about axes in the $xy$-plane. Red sideband pulses $V_s(\theta, \phi) = \bigoplus_{n=1}^\infty Q_n(\theta,\phi)$ can be decomposed into $SU(2)$ rotations, $Q_n(\theta,\phi)$ which are mapped to a $SO(3)$ rotation $R_{\vec{\phi}}(\sqrt{n}\theta)$ where $\vec{\phi} = (\cos{\phi}, \sin{\phi}, 0)$ lies in the $xy$-plane. We decompose $R_{\vec{\mu}_\perp}(\pi)$ into $R_{\vec{a}}(\theta_a) R_{\vec{b}}(\theta_b)$ where $\vec{a}, \vec{b}$ lie in the $xy$-plane (see Appendix \ref{app:decomp}). Then, the inverse mapping $R^{-1}:R_{\vec{a}}(\theta_a) R_{\vec{b}}(\theta_b) \mapsto Q_\mu\left(\frac{\theta_a}{\sqrt{\mu}}, \phi_a\right) Q_\mu\left(\frac{\theta_b}{\sqrt{\mu}}, \phi_b\right)$ specifies the necessary red sideband pulses. Thus, setting $C = V_s\left(\frac{\theta_a}{\sqrt{\mu}}, \phi_a\right) V_s\left(\frac{\theta_b}{\sqrt{\mu}}, \phi_b\right)$, we find
\begin{equation}
        U_{j+1} = U_j C U_j C^\dagger = \bigoplus_{\mu_i \in M_{j+1}}I_{\mu_i}  \oplus \Omega_k^{(j+1)} \bigoplus_{m \in M \setminus M_{j+1}}\Omega_{m}^{(j+1)},
        \label{jth-step}
\end{equation}
where $I_{\mu_i}$ are unaffected by the action of $C$ and $\mu = \mu_{j+1}$ is added to set of cleaned subspaces $M_j$ in Eq. \eqref{uj} to give $M_{j+1}$ in Eq. \eqref{jth-step}. By repeating this procedure $(n-3)$ times built on the base case, we can clean all the subspaces of the $d$-QO-qudit except the $k$-th which we shall deal with in the final step below.

\begin{figure}[ht]
\centering
\begin{subfigure}{\columnwidth}
\includegraphics[width = \columnwidth]{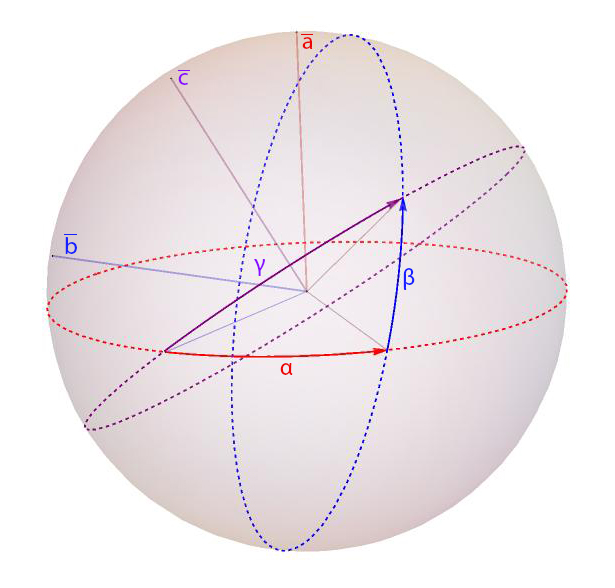}
\caption{}
\label{fig:left}
\end{subfigure}
\begin{subfigure}{\columnwidth}
\includegraphics[width =\columnwidth]{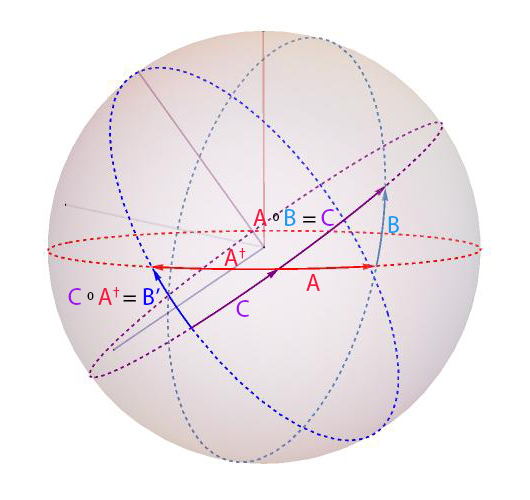}
\caption{}
\label{fig:right}
\end{subfigure}
\caption{(a) Composition of rotations $R_{\vec{a}}(\alpha)R_{\vec{b}}(\beta) = R_{\vec{c}}(\gamma)$ and (b) conjugation of a rotation $B$ by $A$ produces $B'$ which has the same angle as $B$ due to the congruent triangles formed by $A B C$ and $A^\dagger B' C$.}
\label{fig:combined}
\end{figure}

\textbf{Final step} When $j = n-1$, we have cleaned all but the $k$-th subspaces in the $d$-QO-qudit. To transform $\Omega_k^{(n-1)}$ into $\Sigma_k$, we examine the problem in SO(3) taking $R(\Omega_k^{(n-1)}) = R_{\vec{k}}(\theta_k)$ and $R(X) = R_x(\pi)$. WLOG, we only consider $\Sigma_k = X$ since Y belongs to the same conjugacy class and $X^2 = -I$ so we can simply perform our construction for X twice to achieve $-I$. We once again use conjugation to maintain the cleaned subspaces while taking advantage of the fact that the conjugacy classes of $SO(3)$ each consist of all rotations by the same angle, $C(\theta) = \{R_{\vec{r}} (\theta) | \forall \vec{r} \in \mathbb{R}^3\}$, demonstrated in Fig. \ref{fig:right}. 
Using a pair of conjugations
\begin{align}
    &\big[ R_{\vec{r}_1}(\theta_1) R_{\vec{k}}^{l}(\theta_k) R_{\vec{r}_1}(-\theta_1) \big] \big[ R_{\vec{r}_2}(\theta_2) R_{\vec{k}}(\theta_k) R_{\vec{r}_2}(-\theta_2) \big] \nonumber \\
    =& R_{\vec{k}_1}(l\theta_k) R_{\vec{k}_2}(\theta_k)
\end{align}
we can rephrase the problem as finding $\vec{k}_1, \vec{k}_2$ such that $R_{\vec{k}_1}(l\theta_k) R_{\vec{k}_2}(\theta_k) = R_x(\pi)$, given $l = \lceil\frac{\pi}{\theta_k}\rceil-1$ to guarantee rotation by an angle close to $\pi$. We can always find satisfactory $\vec{k}_1, \vec{k}_2$ because the triangle inequality on the sphere states that a composition of two rotations $R_{\vec{a}}(\alpha)R_{\vec{b}}(\beta) = R_{\vec{c}}(\gamma)$ results in a rotation by an angle less than the sum of angles, $|\gamma| \leq |\alpha + \beta|$ as shown in Fig. \ref{fig:left}. So, by choosing $l$ we guarantee that we can achieve an angle of at least $\pi$, and with judicious choice of $\vec{k}_1, \vec{k}_2$ we can achieve a rotation by $\pi$ about any axis. To find the necessary red sideband pulses, we decompose $R_{\vec{r}_1}(\theta_1)$ and $ R_{\vec{r}_2}(\theta_2)$ as before and use the inverse mapping, $R_{\vec{r}_1}(\theta_1) \mapsto Q_k\left(\frac{\theta_{a,1}}{\sqrt{k}}, \phi_{a,1}\right) Q_k\left(\frac{\theta_{b,1}}{\sqrt{k}}, \phi_{b,1}\right)$ and $R_{\vec{r}_2}\left(\theta_2\right) \mapsto Q_k\left(\frac{\theta_{a,2}}{\sqrt{k}}, \phi_{a,2}\right) Q_k\left(\frac{\theta_{b,2}}{\sqrt{k}}, \phi_{b,2}\right)$. Setting $C_p = V_s\left(\frac{\theta_{a,p}}{\sqrt{\mu}}, \phi_{a,p}\right) V_s\left(\frac{\theta_{b,p}}{\sqrt{\mu}}, \phi_{b,p}\right)$, $p \in \{1, 2\}$ we find
\begin{equation}
        V_n^{(k)} = C_1 U_{n-1}^l C_1^\dagger C_2 U_{n-1} C_2^\dagger = I_1 \oplus I_2 \oplus ... \oplus X_k  \oplus ... \oplus I_n...
        \label{vnk-final}
\end{equation}
Thus, we can construct elementary sQM rotations in a $d$-QO-qudit where $d = 2(n+1)$ for any $n$. 
%

\subsubsection{Bounds on the number of pulses required} \label{sec:bound}

Now that we have shown we can create the elementary sQM rotations, we would like to bound the number of red sideband pulses needed to create $V^{(k)}_n$ in this section. One intuition is that the number of pulses required will increase at least exponentially as $n$ increases, since each step in the recursion will cost at least a constant number of pulses due to the conjugation procedure. We shall analyze this more precisely in the following.

To create the identity on any set of $(n-1)$ subspaces in the sQM of a $d$-QO-qudit, we only need $\sim 2^n$ pulses. At the $j$-th recursive step we use $|U_{j+1}| = 4 + 2*|U_j|$ pulses, where $|U_j|$ denotes the number of pulses in the sequence. We can see this by examining Eq. \eqref{jth-step} which uses $|U_j|$ twice and both $C, C^\dagger$ are made up of two red sideband pulses by definition. We begin with $2^2$ pulses at the base step $U_2$, and at each step $j$ we have $|U_j| = \sum_{d=2}^j 2^d = 2^{j+1}-4$. Thus, to clean $j=n-1$ subspaces to identity, we need $(2^n-4)$ pulses. In the final step, a total of $8 + (l+1)*|U_{n-1}| = (l+1)(2^n - 4) + 8$ is required to construct $V_n^{(k)}$ in Eq. \eqref{vnk-final}.

In order to bound $l$, we must understand how the base step and each recursive step $j$ transforms $\Omega_k^{(j)}$ into $\Omega_k^{(j+1)}$. For notation clarity, $(U_j)_m = \Omega_m^{(j)}$ indicates the $SU(2)$ rotation upon the $m$-th subspace in $U_j$ (as defined in Eqs. \eqref{u2}, \eqref{uj}). $\Omega_m^{(j)}$  maps to the $SO(3)$ rotation $R_{\vec{m}_j}\left(\theta_m^{(j)}\right)$. We take the $C, \mu, U_j, U_{j+1}$ as defined in the recursive step above. In the following we will show the conditions such that $\theta_k^{(j)} \geq \frac{\pi}{t}$ for all $j < n$ for some $t > 2$ and $t \in \mathbb{Z}$. 

In the base step, we can easily calculate the rotation angle of $\Omega_k^{(2)} = R_{\vec{k}_2}\left(\theta_k^{(2)}\right)$ from Eq. \eqref{u2}:
\begin{align}
    \theta_k^{(2)} =& 2 \cos^{-1} \bigg\{ \frac{1}{2}\Big[1 + \cos \big(2 \pi \sqrt{\frac{k}{\mu_1}} \big) \nonumber \\
    &- 2 \cos \big( \pi \sqrt{\frac{k}{\mu_2}} \big) \sin^2 \big( \pi \sqrt{\frac{k}{\mu_1}} \big) \Big] \bigg\}.
    \label{base-case-k-angle}
\end{align}
We can always choose $\mu_1, \mu_2 < k$ such that $ \theta_k^{(2)} \geq \frac{\pi}{2}$. For $k \sim 1$, this can be easily calculated explicitly. For general $k$, if we simply take $\mu_1$ as the number closest to $\frac{k}{2}$ that abides by the base case constraint, then Eq. \eqref{base-case-k-angle} is approximately $\theta_k^{(2)} = 2 \cos^{-1}\left[\frac{1}{2}\left(1 + \cos\left(2 \sqrt{2} \pi \right) - 2 \cos\left( \pi \sqrt{\frac{k}{\mu_2}} \right) \sin^2 \left( \sqrt{2} \pi  \right) \right) \right]$. Thus, for $\theta_k^{(2)} \geq \frac{\pi}{2}$, we obtain a periodic condition $ 0.18 <\sqrt{\frac{k}{\mu_2}} - 2 z< 0.74$, $z \in \mathbb{Z}$, which can be easily fulfilled for a range of $\mu_2 < k$. 

We will show the conditions such that $\theta_k^{(j+1)} \geq \frac{\pi}{t}$ if we have $\theta_k^{(j)} \geq \frac{\pi}{t}$ at each recursive step.

Recall, at the $j$-th step we perform $U_{j+1} = U_j C U_j C^\dagger$ where $C =  V_s\left(\frac{\theta_a}{\sqrt{\mu}}, \phi_a\right) V_s\left(\frac{\theta_b}{\sqrt{\mu}}, \phi_b\right)$. Restricting our attention to the $k$-th subspace, we examine $C$ and its action upon $\Omega_k^{(j)} = R_{\vec{k}_j}\left(\theta_k^{(j)}\right)$ in the $SO(3)$ picture:
\begin{align}
       (C)_k &\mapsto R_{\vec{a}} (\theta_a \sqrt{\frac{k}{\mu}} ) R_{\vec{b}} (\theta_b \sqrt{\frac{k}{\mu}} ) \nonumber \\
       &:= R_{\vec{r}_{ab}}(\theta_{ab}), \\
       (C U_j C^\dagger)_k &\mapsto  R_{\vec{r}_{ab}} (\theta_{ab} ) R_{\vec{k}_j} (\theta_k^{(j)} ) R_{\vec{r}_{ab}} (-\theta_{ab} ) \nonumber \\
       &:= R_{\vec{k}_{ab}} (\theta_k^{(j)} ),
\end{align}
where
\begin{align}
       \vec{k}_j \cdot \vec{k}_{ab} =& (\vec{k}_j \cdot \vec{r}_{ab} )^2 +  [1 - (\vec{k}_j \cdot \vec{r}_{ab})^2 ] \cos{\theta_{ab}} \label{k-ab} \\
       \theta_{ab} =& 2 \cos^{-1} \bigg\{ \frac{1}{2} \Big[ \cos \big( \frac{\theta_a}{2} \sqrt{\frac{k}{\mu}} \big) \cos \big( \frac{\theta_b}{2} \sqrt{\frac{k}{\mu}} \big) \nonumber  \\
       & - (\vec{a} \cdot \vec{b} ) \sin \big( \frac{\theta_a}{2} \sqrt{\frac{k}{\mu}} \big) \sin\big( \frac{\theta_b}{2} \sqrt{\frac{k}{\mu}} \big) \Big] \bigg\} \label{theta-ab}
\end{align}
Recall, $\vec{a}, \vec{b}, \theta_a, \theta_b$ are found by decomposing $R_{\vec{\mu}_\perp}(\pi) = R_{\vec{a}}(\theta_a) R_{\vec{b}}(\theta_b)$ into two rotations about axes in the $xy$-plane. Using these definitions we can write the outcome of $\Omega_k^{(j+1)} \mapsto R_{\vec{k}_{j+1}}\left(\theta_k^{(j+1)}\right) = R_{\vec{k}_j}\left(\theta_k^{(j)}\right)R_{\vec{k}_{ab}}\left(\theta_k^{(j)}\right)$ in $SO(3)$. We write explicitly our condition for $\theta_k^{(j+1)}$:
\begin{align}
        \theta_k^{(j+1)}
        =& 2 \cos^{-1}\left[\frac{1}{2}\left( 1 - \vec{k}_j \cdot \vec{k}_{ab} + \left(1 + \vec{k}_j \cdot \vec{k}_{ab}\right) \cos{\theta_k^{(j)}}   \right)\right]  \nonumber \\
        \geq& \frac{\pi}{t}
        \label{theta-j+1}
\end{align}
Substituting $\vec{k}_j \cdot \vec{k}_{ab}$ with Eq. \eqref{k-ab} into Eq. \eqref{theta-j+1}, we can rewrite the condition as a constraint on $\theta_{ab}$ or on $\vec{r}_{ab}$:
\begin{equation}
\begin{split}
    |\vec{k}_j \cdot \vec{r}_{ab}| &\geq \sqrt{\frac{\cos{\frac{\pi}{2 t}}-1}{\cos{\theta_k^{(j)}}-1}} \\
    \cos{}\theta_{ab} &\geq \frac{1- 2 \cos{\frac{\pi}{2 t}}-\cos{\theta_k}}{1-\cos{\theta_k^{(j)}}}
\end{split}
\label{constraint-final}
\end{equation}
This indicates that we need $R_{\vec{r}_{ab}}(\theta_{ab})$ to not be close to $R_{\vec{k}_{j, \perp}}(\pi)$ where $\vec{k}_{j, \perp} \perp \vec{k}_j$. When $t \sim 2^4$ and $|\theta_k^{(j)}| \geq \frac{2 \pi}{t}$, this condition is quite flexible and restricts $\cos{\theta_{ab}} > -0.873$ or $|\vec{k}_j \cdot \vec{r}_{ab}|  > 0.26$. And when $|\theta_k^{(j)}| \geq \frac{4 \pi}{t}$, $\cos{\theta_{ab}} > -0.967$ or $|\vec{k}_j \cdot \vec{r}_{ab}|  > 0.128$. At worst, when $|\theta_k^{(j)}| = \frac{\pi}{t}$, the restriction is that $\cos{\theta_{ab}} > -\frac{1}{2}$ or $|\vec{k}_j \cdot \vec{r}_{ab}|  > \frac{1}{2}$. When $\mu \ll k$ or $\mu \gg k$, the conjugation rotations $C$ affect the $k$-th and $\mu$-th subspaces very differently due to the rotation $(C)_\mu$ and $(C)_k$ being composed of rotations about the same axes, but by very different angles, indicated by the ratio of $\frac{k}{\mu}$. Thus, by changing how we decompose $C$ into rotations in the $xy$-plane, we can ensure the flexibility of $R_{\vec{r}_{ab}}(\theta_{ab})$. When $\mu \sim k$, specifically, when $\sqrt{\frac{k}{\mu}} \sim 1$, $\theta_{ab}$ will be close to $\pi$ because $(C)_{\mu}$ is a rotation by $\pi$. Thus, the only way to ensure $\theta_k^{(j)}$ is maintained, is to clean such $\mu$ when $\vec{\mu} \cdot \vec{k}_j \lesssim \frac{1}{2}$, i.e. $(U_j)_\mu$ and $(U_j)_k$ are rotations about different enough axes. Also, note that $\frac{1}{2}$ is a worst case scenario, when $(U_j)_k$ is a rotation by precisely $\frac{\pi}{t}$, and from our discussion of Eq. \eqref{constraint-final} this constraint becomes relaxed quickly. We also note that when $\mu \gg k$, the effect of the conjugation sequence upon the $k$-th subspace is negligible and so $\vec{k}_{ab} \sim \vec{k}_j \implies \theta_k^{(j+1)} \sim 2*\theta_k^{(j)}$.

In order to roughly bound $l$ for asymptotically large $n$, we consider two different cases: $k \ll n$ or $k \sim n$. When $k$ is much less than $n \gg 1$, $l$ is bounded by $4$ by choosing $t = 4$. This is because the majority of recursive steps involve $\mu \gg k$, thus we have many choices in the order of the cleaning to almost double the angle on the $k$-th subspace at a given step. Thus, whenever $\theta_k^{(j)} \sim \frac{\pi}{t}$, we can clean a subspace such that $\theta_k^{(j)} \sim \frac{2 \pi}{t}$, which provides an ideal constraint from Eq. \eqref{constraint-final} for the next recursive step. And so we can improve the angle on the $k$-th space to ensure cleaning subspaces near $k$ can be done optimally. When $k \gg 1$, we have the opposite situation because for large  $k$, $\Delta_k = \sqrt{k+1} - \sqrt{k} = \frac{1}{2\sqrt{k}}$. Thus, on the order of $\sqrt{k}$ pulses are needed to separate the rotations on the $k$ and $(k+1)$-th subspaces. However, by cleaning the subspaces outside the range $k \pm \sqrt{k}$, we are in effect separating the subspaces near $k$ from each other, albeit slower than directly applied pulses. This is because $(C)_k, (C)_{k+1}$ perform almost the same rotation and differ slightly in the angle by on the order $\frac{1}{\sqrt{k}}$. Thus, at every recursive step, we build up a difference in the axis of rotation between the $k$ and $k+1$ subspace of about $\frac{1}{k}$. So after $\sim \frac{k}{2}$ recursive steps, we build up a difference of $\sim \frac{1}{2}$, the necessary difference for the worst case scenario in maintaining $\theta_k^{(j+1)}$ when $\theta_k^{(j)} = \frac{\pi}{t}$. Of course, we cannot guarantee that all subspaces within $ k \pm \sqrt{k}$ can be optimally separated from $k$ at the same time, and so $k \leq t \leq k^{2\sqrt{k}}$ to capture the worst case scenarios of cleaning between one and $2\sqrt{k}$ subspaces while their axes of rotation are parallel to the $k$-th subspace's axis of rotation at each recursive step. Therefore, as it is highly unlikely that the $2\sqrt{k}$ subspaces will all be exactly parallel to $k$ and given our freedom to clean subspaces, the majority of which are not near the $k$-th, in any order, we generally bound $l$ by $k^2$. 

In summary, a total of $(l+2)(2^n - 4) + 8$ sideband pulses are needed to construct any clean elementary SU(2) rotation in sQM in Eq. \eqref{vnk-final}, where $l$ is roughly $O(k^2)$ and $k$ is the subspace index ($1 \le k \le n$). We know the decomposition of an arbitrary $d\times d$ unitary operation needs at most $d(d-1)/2$ two-level unitaries \cite{nielsen2010quantum}, and each two-level unitary can be further constructed from roughly $d$ clean elementary SU(2) rotations in sQM as described in Sec. \ref{sec:cleanElemSU2ToUniversality} (Eqs. \eqref{sQM-clean-ele}\eqref{sQM-vecZ}\eqref{sQM-final-refocus}\eqref{cQM-clean-ele}\eqref{cQM-final-refocus}). Thus, the total number of control pulses needed for an arbitrary $d\times d$ unitary operation is $O(d^5 2^{d/2})$, or equivalently $O(n^5 2^n)$ in terms of $n$ since $d = 2(n+1)$. Note that this scaling comes from a worst case estimation, and for typical unitaries one might expect a better scaling.

\section{Algorithms and Examples \label{sec:examples}}

We have also implemented an algorithm according to 
the above constructive proof to produce composite pulse sequences that can realize arbitrary elementary clean $SU(2)$ unitary operations. We describe our algorithm in detail in Sec. \ref{sec:6d-example}, where an example construction of clean elementary Pauli $X$ gate for an 8-QO-qudit (3 subspaces in sQM and 4 subspaces in cQM) is given. We then present in Sec. \ref{sec:num-scaling} more extensive numerical results on the construction of clean elementary Pauli $X$ gates in $d$-QO-qudits ($d = 2(n+1)$) for all $n \le 22$ and demonstrate the agreement between numerical and theoretical bounds on the number of sideband pulses.

\subsection{Details of the Algorithm and Example Construction of Pauli $X$ Gate on a 8-QO-Qudit}
\label{sec:6d-example}

Following is an algorithm for constructing a gate with identity operations $I$ on all of the subspaces apart from subspace $k$, which has an $X$ or $Y$ gate (i.e., $I_1 \oplus \cdots \oplus \Sigma_k \oplus \cdots \oplus I_n$, where $\Sigma_k \in \{X, Y\}$).  This algorithm parallels the recursive procedure described in Sec. \ref{sec:inductiveProof}; however, we do not implement the strategies described in Sec. \ref{sec:bound} but rather rely on empirical optimization over the four degrees of freedom described at the end of \ref{sec:6d-example} to achieve efficient pulse sequences.

Note that we adopt the convention of using the $SO(3)$ representation in which a rotation about unit vector $\hat{r}$ by angle $\theta$ is represented by $R_{\hat{r}}(\theta)$ throughout the description of the algorithm.  Once our sequence involves only rotations about unit vectors in the $xy$-plane, we can convert each rotation into the $SU(2)$ representation $V_s(\theta, \phi)$ for red sideband pulses.

\begin{enumerate}
    \item Construct an ordered pulse sequence $S_I$ (order goes from left to right) that results in the operation $I_1 \oplus \cdots \oplus \Omega_k \oplus \cdots \oplus I_n$.
    \begin{enumerate}
        \item Select distinct $\mu_1, \mu_2 \in \mathbb{Z}$ with $1 \leq \mu_1, \mu_2 \leq n$ such that $\sqrt{\frac{k}{\mu_1}}, \sqrt{\frac{k}{\mu_2}} \not\in \mathbb{Z}$.
        \item Initialize $S_I$ to be the four-pulse sequence
        \begin{equation}
            S_I \leftarrow \left\{ R_{\hat{y}}(\frac{\pi}{\sqrt{\mu_1}}), R_{\hat{x}}(\frac{2 \pi}{\sqrt{\mu_2}}), R_{\hat{y}}(\frac{\pi}{\sqrt{\mu_1}}), R_{\hat{x}}(-\frac{2 \pi}{\sqrt{\mu_2}}) \right\}
        \end{equation}
        that leaves $I_{\mu_1}$ and $I_{\mu_2}$ on subspaces $\mu_1$ and $\mu_2$ unchanged while rotating subspace $k$.
        \item For each subspace $m$ apart from $k$, $\mu_1$, and $\mu_2$ (i.e., for each $m \in \mathbb{Z}$ with $1 \leq m \leq n$, excluding $k$, $\mu_1$, and $\mu_2$), suppose that the operation on this subspace is $\Omega_m$ after the application of pulse sequence $S_I$.  If $\Omega_m = I_m$, continue to the next such $m$.  Otherwise, let $\hat{r}_m$ and $\theta_m$ be the axis and angle of the rotation induced by $\Omega_m$, respectively, and then set
        \begin{equation}
            S_I \leftarrow \left\{S_I, R_{\hat{r}^\perp_m} (-\frac{\pi}{\sqrt{m}} ), S_I, R_{\hat{r}^\perp_m} (\frac{\pi}{\sqrt{m}} )\right\},
        \end{equation}
        where $\hat{r}^\perp_m$ is any unit vector perpendicular to $\hat{r}_m$.
        
        Continue to the next such $m$ and repeat, using the operation $\Omega_m$ on subspace $m$ following the application of the updated sequence $S_I$.
    \end{enumerate}
    \item Use pulse sequence $S_I$ to place an $X$ or $Y$ gate on subspace $k$.
    \begin{enumerate}
        \item Letting $\theta_k$ be the angle corresponding to rotation $\Omega_k$ and $\hat{r}_k$ the corresponding axis, calculate $l = \lceil \frac{\pi}{\theta_k} \rceil - 1$, and define the components of unit vectors $\hat{\alpha}$ and $\hat{\beta}$ as follows.  If the desired gate $\Sigma_k$ is $X$, then we have
        \begin{align}
            \alpha_1 &= \cos{\frac{l \theta_k}{2}} \csc{\frac{\theta_k}{2}}\\
            \alpha_2 &= \csc{\frac{\theta_k}{2}} \sqrt{-\cos^2{\frac{\theta_k}{2}} + (1 - \beta_2^2) \sin^2{\frac{l \theta_k}{2}}}\\
            \alpha_3 &= -\beta_2 \csc{\frac{\theta_k}{2}} \sin{\frac{l \theta_k}{2}}\\
            \beta_1 &= \cos{\frac{\theta_k}{2}} \csc{\frac{l \theta_k}{2}}\\
            \beta_3 &= \csc{\frac{l \theta_k}{2}} \sqrt{-\cos^2{\frac{\theta_k}{2}} + (1 - \beta_2^2) \sin^2{\frac{l \theta_k}{2}}},
        \end{align}
        where $\beta_2$ is a free real parameter with
        \begin{equation}
            |\beta_2| \leq \sqrt{1 - \frac{\cos^2{\frac{\theta_k}{2}}}{\sin^2{\frac{l \theta_k}{2}}}}
        \end{equation}
        such that all of the components of $\hat{\alpha}$ and $\hat{\beta}$ are real.  Note that this range is guaranteed to be nonempty because it follows from the definition of $l$ by $l = \lceil \frac{\pi}{\theta_k} \rceil - 1$ that
        \begin{equation}
            \sin^2{\frac{l \theta_k}{2}} \geq \cos^2{\frac{\theta_k}{2}}.
        \end{equation}
        Similarly, if the desired gate $\Sigma_k$ is $Y$, then we have
        \begin{align}
            \alpha_1 &= -\csc{\frac{\theta_k}{2}} \sqrt{-\cos^2{\frac{\theta_k}{2}} + (1 - \beta_1^2) \sin^2{\frac{l \theta_k}{2}}}\\
            \alpha_2 &= \cos{\frac{l \theta_k}{2}} \csc{\frac{\theta_k}{2}}\\
            \alpha_3 &= \beta_1 \csc{\frac{\theta_k}{2}} \sin{\frac{l \theta_k}{2}}\\
            \beta_2 &= \cos{\frac{\theta_k}{2}} \csc{\frac{l \theta_k}{2}}\\
            \beta_3 &= \csc{\frac{l \theta_k}{2}} \sqrt{-\cos^2{\frac{\theta_k}{2}} + (1 - \beta_1^2) \sin^2{\frac{l \theta_k}{2}}},
        \end{align}
        where $\beta_1$ is a free real parameter with
        \begin{equation}
            |\beta_1| \leq \sqrt{1 - \frac{\cos^2{\frac{\theta_k}{2}}}{\sin^2{\frac{l \theta_k}{2}}}}
        \end{equation}
        such that all of the components of $\hat{\alpha}$ and $\hat{\beta}$ are real, where this range is nonempty for the same reason as that provided above.
        
        \item For $\hat{\beta}$, calculate
        \begin{align}
            \hat{r}_\beta &= \frac{\hat{r}_k \times \hat{\beta}}{\left\lVert\hat{r}_k \times \hat{\beta}\right\rVert}, \\
            \theta_\beta &= \cos^{-1}\left(\hat{r}_k \cdot \hat{\beta}\right),
        \end{align}
        and calculate $\hat{r}_\alpha$ and $\theta_\alpha$ similarly for $\hat{\alpha}$.
        Finally, the desired sequence is now
        \begin{align}
            S_I \leftarrow \bigg\{ &R_{\hat{r}_\beta} (-\frac{\theta_\beta}{\sqrt{k}} ), S_I^l, R_{\hat{r}_\beta} (\frac{\theta_\beta}{\sqrt{k}} ), \nonumber \\
            &R_{\hat{r}_\alpha} (-\frac{\theta_\alpha}{\sqrt{k}} ), S_I, R_{\hat{r}_\alpha} (\frac{\theta_\alpha}{\sqrt{k}} ) \bigg\},
        \end{align}
        where $S_I^l$ denotes the $l$-fold repetition of the sequence $S_I$.
    \end{enumerate}
\end{enumerate}
To obtain a sequence with $\Sigma_k = -I$, the algorithm can be carried out for $\Sigma_k = X$ and the resulting sequence repeated, since throughout this work we define $X = i \sigma_x$, and $(i \sigma_x)^2 = -I$.

Note that the rotation unit vectors $\hat{r}^\perp_m$, $\hat{r}_\beta$, and $\hat{r}_\alpha$ need not be in the $xy$-plane, and so the rotations about these vectors might not be achievable by a single rotation with axis in the $xy$-plane.  However, these rotations---and, more generally, any rotation $R_{\hat{r}}(\theta)$---can also be algorithmically decomposed into an equivalent pair of $xy$-plane rotations according to the subroutine in Appendix \ref{app:subroutine-xy-decomp}.  With this subroutine, any rotations about axes not in the $xy$-plane are converted into sequences of two rotations about axes in the $xy$-plane, thus ensuring that all rotations in our final sequence are about axes in the $xy$-plane.  Finally, these rotations about axes in the $xy$-plane in $SO(3)$ notation of the form $R_{\hat{r}}(\theta)$ with $\hat{r} = \langle r_1, r_2, 0 \rangle$ can be converted into the form of red sideband pulses $V_s(\theta, \phi)$ by the equivalence
\begin{equation}
    R_{\hat{r}}(\theta) \leftrightarrow V_s(\theta, \atantwo{(r_2, r_1)}),
\end{equation}
where $\atantwo$ denotes the standard 2-argument arctangent function.

Finally, it is worth noting that there are four primary degrees of freedom in this algorithm---namely,
\begin{enumerate}[i.]
    \item the choice of $\mu_1, \mu_2$ for the initial four-pulse $S_I$ sequence;
    \item the order in which the operators on the remaining subspaces are converted to the identity;
    \item the axis $\hat{r}_m^\perp$ of the rotation used to conjugate $S_I$ when converting the operator on subspace $m$ to the identity; and
    \item the angle $\phi$ in each decomposition of a rotation about an axis $\hat{r}$ not in the $xy$-plane into a sequence of two rotations about axes $\hat{r}_1$ and $\hat{r}_2$ in the $xy$-plane.
\end{enumerate}
These degrees of freedom can be chosen accordingly to empirically minimize the number of pulses in the sequence by ensuring that $\theta_k$ is near $\pi$ after Step 1 of the algorithm, which in turn minimizes $l$ and thus minimizes repetition of the $S_I$ sequence.

This algorithm has been written in Python, and the code \cite{mintzer2021repo} has been used to generate sequences of red sideband pulses of the the form $V_s(\theta, \phi)$ that produce the gates $I_1 \oplus I_2 \oplus X_3$ and $I_1 \oplus I_2 \oplus Y_3$.  The resulting parameters for the sequences are provided in Table \ref{tab:pulse_seq_iix_iiy}.

\begin{table}[th]
    \centering
    \begin{tabular}{cccc} \toprule
        \multicolumn{2}{c}{$I_1 \oplus I_2 \oplus X_3$} & \multicolumn{2}{c}{$I_1 \oplus I_2 \oplus Y_3$} \\
        $\theta$ & $\phi$ & $\theta$ & $\phi$ \\ \midrule
        $-1.0956$ & $-2.8651$ & $-1.3098$ & $-2.7835$ \\
        $-\pi / \sqrt{3}$ & $0.4867$ & $-\pi / \sqrt{3}$ & $0.4210$ \\
        $\sqrt{2} \pi $ & $\pi / 2$ & $\sqrt{2} \pi $ & $\pi / 2$ \\
        $\pi$ & $0$ & $\pi$ & $0$ \\
        $\sqrt{2} \pi $ & $\pi / 2$ & $\sqrt{2} \pi $ & $\pi / 2$ \\
        $-\pi$ & $0$ & $-\pi$ & $0$ \\
        $\pi / \sqrt{3}$ & $0.4867$ & $\pi / \sqrt{3}$ & $0.4210$ \\
        $1.0956$ & $-2.8651$ & $1.3098$ & $-2.7835$ \\
        $-1.8073$ & $-0.3267$ & $-1.4071$ & $-2.6391$ \\
        $-2.0499$ & $-\pi$ & $-1.2674$ & $0$ \\
        $\sqrt{2} \pi $ & $\pi / 2$ & $\sqrt{2} \pi $ & $\pi / 2$ \\
        $\pi$ & $0$ & $\pi$ & $0$ \\
        $\sqrt{2} \pi $ & $\pi / 2$ &  $\sqrt{2} \pi $ & $\pi / 2$ \\
        $-\pi$ & $0$ & $-\pi$ & $0$ \\
        $2.0499$ & $-\pi$ & $1.2674$ & $0$ \\
        $1.8073$ & $-0.3267$ & $1.4071$ & $-2.6391$ \\ \bottomrule
    \end{tabular}
    \caption{Pulse sequences for $I_1 \oplus I_2 \oplus X_3$ and $I_1 \oplus I_2 \oplus Y_3$}
    \label{tab:pulse_seq_iix_iiy}
\end{table}

\subsection{Numerical Scaling of the Number of Pulses}
\label{sec:num-scaling}

\begin{figure}[ht!]
    \centering
    \includegraphics[width=\columnwidth]{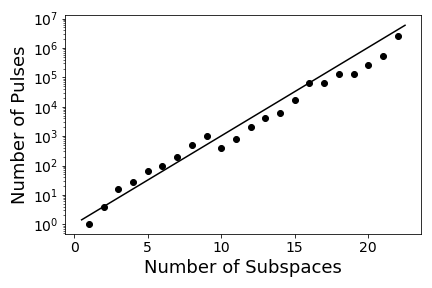}
    \caption{Lengths of pulse sequences for constructions of gates of the form $I_1 \oplus I_2 \oplus \cdots \oplus I_{n-1} \oplus X_n$ up to 22 subspaces, where the solid line indicates the theoretical scaling of $2^n$ for $n$ subspaces. The small deviation of the numerical points from the theoretical bound is due to the degrees of freedom for tuning our algorithm, as detailed in Sec. \ref{sec:6d-example}. }
    \label{fig:pulses_vs_subspaces}
\end{figure}

To support the theoretical bound we derived on the construction in Sec. \ref{sec:bound}, we also use our code to generate pulse sequences to construct gates of the form $I_1 \oplus I_2 \oplus \cdots \oplus I_{n-1} \oplus X_n$ up to the lowest $23$ 
Fock levels of the oscillator, i.e., 22 subspaces in sQM (excluding states $\ket{0,0}$ and $\ket{1,n}$ from the sQM because they undergo trivial transformations in red sideband pulses). We plot the total number of sideband pulses used in each case as a function of $n$, as is shown in Fig. \ref{fig:pulses_vs_subspaces}. The approximate linear dependence in log-scale empirically demonstrates the exponential scaling of the number of pulses with the QO-qudit dimension (or the truncated oscillator dimension), which is consistent with the bound proven in Sec. \ref{sec:bound}. Note the small deviation of the numerical results from exact exponential scaling in Fig. \ref{fig:pulses_vs_subspaces} is a manifestation of the several degrees of freedom mentioned in Sec. \ref{sec:6d-example} that we can tune, which yields a slightly different number of pulses for each $n$. Also note the gates $I_1 \oplus I_2 \oplus \cdots \oplus I_{n-1} \oplus X_n$ synthesized here are all elementary SU(2) rotations in the sQM, so the polynomial prefactor $n^5$ (as described in Sec. \ref{sec:universality} for an arbitrary $n\times n$ unitary) does not apply.

\section{Conclusions \label{sec:conclusion}}

We have shown that an infinite dimensional harmonic oscillator can be truncated to a finite dimensional subspace $(n+1)$ for any $n \in \mathbb{Z}$ by coupling to a single qubit, to form a $d$-dimensional QO-qudit for $d = 2(n+1)$. A recursive construction is given to synthesize arbitrary unitary operations in the QO-qudit using only the red sideband and the carrier pulses \emph{fully analytically}. These control pulses used in our construction are routinely available in a Jaynes-Cummings type interaction as is easily realized in many physical platforms including trapped ions. The ability to synthesize arbitrary unitary operations in a $d$-QO-qudit immediately implies universal control on the corresponding truncated oscillator. At the heart of our construction is the utilization of the naturally repeated pattern in the spectrum of an oscillator. This repeated pattern allows us to recursively clean each two-dimensional subspace by the dynamical decoupling technique. To analyze the scaling of our construction, a bound on the number of sideband pulses required to accomplish the construction is derived. It is shown that the number of sideband pulses scales exponentially as the dimension $d$ of a QO-qudit with a low degree polynomial prefactor depending on $d$.

We believe the exponential scaling derived is optimal and cannot be reduced to a polynomial scaling, if the synthesized unitary is completely arbitrary and is constructed fully analytically as in our work. From the viewpoint of complexity theory, it also indicates the task of closing a low energy subspace in an oscillator and construct an arbitrary unitary operation to \emph{arbitrary precision} is exponentially hard. The exponential scaling of the number of pulse also translates to an exponential scaling of total energy required. Therefore, closing a finite dimensional low energy subspace of the oscillator does {\em not} surmount the unphysically infinite amount of energy requirement on the universal control of an oscillator. However, following \cite{mischuck2013qudit}, numerical optimizations may be used to reduce pulse requirements, and it is likely that significant simplifications can be realized for specific unitaries, especially those with additional structure.

It should be noted that our construction should work for a larger class of Hamiltonians beyond the JC Hamiltonian. One example is for trapped ions beyond the deep Lamb-Dicke regime where the Rabi frequency of the subspaces are proportional to the Legendre polynomial, instead of a simple $\sqrt{n}$ dependence as discussed above. It would be interesting to generalize this to other Hamiltonians where unbounded bosonic systems are utilized for quantum computation, such as transmon+microwave cavity in the superconducting architecture. Moreover, by hybridizing the unitaries constructed under the Fock basis in this work with various continuous-variable type operations, more efficient and powerful operations are likely to arise.

Our discussions presented above assume perfectly isolated qubit-oscillator systems. In practical applications, the qubit-oscillator system may be coupled to an external noisy environment, leading to the presence of quantum noise in QOQ. In such noisy cases, the scheme proposed above cannot be directly applied. However, this can be circumvented by combining our protocol with proper quantum error-correcting codes. For example, depending on the nature of the quantum noise (system-bath coupling), a subspace of our qubit-oscillator system may be identified that is immune to the noise produced by the environment. Therefore, our protocol can be applied to the decoherence-free subspace \cite{lidar1998decoherence} of the qubit-oscillator system. More generally, we may use our protocol to implement arbitrary unitary operations directly on the \textit{logical qubits/qudits} (instead of the Fock levels) defined by a given quantum error-correcting code to get rid of the possible errors induced by coupling to external environment. Any unitary operations on these logical states can be implemented by our protocol, as is guaranteed by the universality proved in Sec. \ref{sec:universality}. 

As a final note, the present work suggests that the combination of qubit+bosonic system may serve as hardware-efficient quantum resources for computational and information storage. Going beyond a single QO-qudit, our constructions may be generalized to include interactions between two QO-qudits, for realizing QO-qudit-based universal quantum computation as was pioneered in Ref. \cite{mischuck2013qudit}.  We hope that analytical approaches, such as those demonstrated here, will lead to further understanding of the algebraic structure of the tensor product of QO-qudits.

\begin{acknowledgments}
The authors thank Steven Girvin and Nathan Wiebe for helpful discussions. 
YL was funded in part by NSF grant PHY-1818914.
YL and GM were supported in part by NTT Research.  GM, JSS, and MTK were supported in part by the Army Research Office, under the CVQC project W911NF-17-1-0481, and the metastable qubits project W911NF-20-1-0037.  The work of YL and ILC on analysis was supported by the U.S. Department of Energy, Office of Science, National Quantum Information Science Research Centers, Co-Design Center for Quantum Advantage, under contract number DE-SC0012704.
\end{acknowledgments}

\appendix
\section{Definition of the Generalized Gell-Mann Matrices \label{app:definitionGM}}

In general, for a $d \times d$ Hermitian matrix $H$, we can find a set of basis matrices, such that $H$ can be written as a linear combination of the basis matrices. Denote $E_{ij}$ as the $d \times d$ matrix with the $(i,j)$-th element being 1 and the rest are zeros. Recall that for the $2\times2$ dimensional case where there are four Pauli matrices, i.e., $I, \sigma_z, \sigma_x, \sigma_y$. In analogy to $\sigma_z$, we can define $(d-1)$ diagonal matrices $M_j^Z$ in the following way
\begin{eqnarray}
    M^Z_{j} = \sqrt{\frac{2}{j(j-1)}} \bigg(\sum_{k=1}^{j-1} E_{kk} - (j-1) E_{jj} \bigg),
\end{eqnarray}
where $ 2 \le j \le d$. Similarly, there are $d(d-1)/2$ real matrices $M_{jk}^{X}$
\begin{align}
    M_{jk}^{X} = E_{jk} + E_{kj}~~ (2 \le j < k \le d)
\end{align}
in analogy to $\sigma_x$, and another $d(d-1)/2$ imaginary matrices $M_{jk}^{Y}$
\begin{align}
    M_{jk}^{Y} = -i E_{jk} + i E_{kj} ~~ (2 \le j < k \le d)
\end{align}
analogous to $\sigma_y$. Simple counting suggests that there are a total of $d^2-1$ such basis matrix (without the identity matrix).
It can be verified that they are traceless and orthonormalized to each other. Moreover, they are closed under the commutation
\begin{align}
    [M_j^Z, M_k^Z] =& 0, \\
    [M_j^Z, M_{lm}^{X}] =& i\sqrt{\frac{2}{j(j-1)}} \bigg(\sum_{k=1}^{j-1} (\delta_{km} M_{kl}^{Y} + \delta_{lk} M_{km}^{Y}) \nonumber \\ 
                         &-(j-1) ( \delta_{mj} M_{jl}^{Y} + \delta_{lj} M_{jm}^{Y})  \bigg), \\
    [M_j^Z, M_{lm}^{Y}] =& i\sqrt{\frac{2}{j(j-1)}} \bigg( \sum_{k=1}^{j-1} (\delta_{mk} M_{kl}^{X} - \delta_{lk} M_{mk}^{X}) \nonumber \\
                         &-(j-1) ( \delta_{mj} M_{jl}^{X} - \delta_{lj} M_{mj}^{X})  \bigg),
\end{align}
\begin{align}
    [M_{jk}^{X}, M_{lm}^{X}] =& i\delta_{kl} M_{jm}^{Y} + i \delta_{jl} M_{km}^{Y} + i \delta_{mj} M_{kl}^{Y} + i\delta_{mk} M_{jl}^{Y}, \\
    [M_{jk}^{Y}, M_{lm}^{Y}] =& -\delta_{kl} M_{jm}^{X} + \delta_{jl} M_{km}^{X} + \delta_{mj} M_{kl}^{X} - \delta_{mk} M_{jl}^{X}, \\
    [M_{jk}^{X}, M_{lm}^{Y}] =& i\delta_{km} M_{jl}^{X} - i \delta_{jl} M_{km}^{X} + i \delta_{mj} M_{kl}^{X} - i\delta_{lk} M_{mj}^{X}.
\end{align}

\section{Decomposing an Arbitrary Rotation into Two Rotations in the Same Plane} 
\label{app:decomp}
A well-known fact is any SO(3) rotation, $R_{\vec{r}}(\theta)$ can be decomposed into two reflections about planes, $p_1, p_2$ intersecting at an angle $\theta/2$ such that $\vec{r} = p_1 \cap p_2$.  
\begin{equation}
    R_{\vec{r}}(\theta) = S(p_1) \circ S(p_2)
    \label{rot-refl}
\end{equation} 
Composing a reflection with itself produces the identity, thus we can insert a reflection about plane $p$ into Eq. \eqref{rot-refl} and rewrite $R_{\vec{r}}(\theta) = S(p_1) \circ S(p) \circ S(p) \circ S(p_2)$. We would like to decompose arbitrary rotations into two red sideband pulses, each of which is restricted to rotation about axis in the xy plane, so we choose $p$ to be the $xy$ plane. Then, we find $R_{\vec{r}}(\theta) = R_{\vec{r}_1}(\theta_1) R_{\vec{r}_2}(\theta_2)$, with $\vec{r}_1 = p_1 \cap p$ and $\vec{r}_2 = p \cap p_2$, and $\theta_i/2 = |\angle (p_i, p)|$, the angle between the planes. We can always choose $p_1, p_2$ to be different from the xy plane, so we are guaranteed to find a decomposition into two red sideband pulses. 

\section{Subroutine on Decomposition of Rotations in to the xy-plane} \label{app:subroutine-xy-decomp}

Here is a detailed subroutine to decompose any rotation $R_{\hat{r}}(\theta)$ into rotations with rotation axes lying in the $xy$-plane:
\begin{enumerate}
    \item If $\hat{r}$ is already in the $xy$-plane (i.e., its $z$-component is equal to zero), then there is no need for decomposition, and the rotation can be made directly.
    \item If $\hat{r}$ is not in the $xy$-plane, then find a unit vector $\hat{r}_1^\perp$ that is perpendicular to $\hat{r} = \langle r_1, r_2, r_3 \rangle$.  Because the $z$-component $r_3$ is nonzero, such a vector is given by 
    
    \begin{equation}
        \hat{r}_1^\perp = \frac{1}{\sqrt{r_1^2 + r_3^2}} \langle -r_3, 0, r_1 \rangle.
    \end{equation}
    
    \item Find a second unit vector $\hat{r}_2^\perp$ that is perpendicular to both $\hat{r}$ and $\hat{r}_1^\perp$ by taking
    
    \begin{equation}
        \vec{r}_2^\perp = \frac{\hat{r} \times \hat{r}_1^\perp}{\lVert \hat{r} \times \hat{r}_1^\perp \rVert}.
    \end{equation}
    
    \item For any value of $\phi \in \mathbb{R}$, $0 \leq \phi < 2 \pi$, take the unit normal vectors
    
    \begin{align}
        \hat{n}_1 &= \hat{r}_1^\perp \cos{\left(\phi\right)} + \hat{r}_2^\perp \sin{\left(\phi\right)} \\
        \hat{n}_2 &= \hat{r}_1^\perp \cos{\left(\phi + \frac{\theta}{2}\right)} + \hat{r}_2^\perp \sin{\left(\phi + \frac{\theta}{2}\right)},
    \end{align}
    
    where a reflection through the plane with normal vector $\hat{n}_1$ followed by a reflection through the plane with normal vector $\hat{n}_2$ is equivalent to a rotation about $\hat{r}$ by an angle of $\theta$.
    
    \item Next, as per the discussion in Appendix \ref{app:decomp}, two reflections about the $xy$-plane (which together are equivalent to the identity transformation) are inserted, and this sequence is recomposed into a sequence of two rotations about unit vectors $\hat{r}_1$ and $\hat{r}_2$ in the $xy$-plane.
    
    In particular, define $\hat{z} = \langle 0, 0, 1 \rangle$ to be the unit normal vector to the $xy$-plane, and take
    
    \begin{align}
        \hat{r}_1 &= \frac{\hat{n}_1 \times \hat{z}}{\lVert \hat{n}_1 \times \hat{z} \rVert} \\
        \theta_1 &= 2 \cos^{-1}{\left( \hat{n}_1 \cdot \hat{z} \right)} \\
        \hat{r}_2 &= \frac{\hat{z} \times \hat{n}_2}{\lVert \hat{z} \times \hat{n}_2 \rVert} \\
        \theta_2 &= 2 \cos^{-1}{\left( \hat{z} \cdot \hat{n}_2 \right)}.
    \end{align}
    
    It is guaranteed that $\hat{r}_1$ and $\hat{r}_2$ lie in the $xy$-plane because they must, by definition, be perpendicular to $\hat{z}$.  Moreover, these choices give the equivalence 
    
    \begin{equation}
        R_{\hat{r}}(\theta) \leftrightarrow \{R_{\hat{r}_1}(\theta_1), R_{\hat{r}_2}(\theta_2)\}
    \end{equation}
    
    of the sequence of rotations $\{R_{\hat{r}_1}(\theta_1), R_{\hat{r}_2}(\theta_2)\}$ about axes in the $xy$-plane with the original rotation $R_{\hat{r}}(\theta)$, as desired.
\end{enumerate}

\bibliography{ref}

\begin{thebibliography}{36}%
\makeatletter
\providecommand \@ifxundefined [1]{%
 \@ifx{#1\undefined}
}%
\providecommand \@ifnum [1]{%
 \ifnum #1\expandafter \@firstoftwo
 \else \expandafter \@secondoftwo
 \fi
}%
\providecommand \@ifx [1]{%
 \ifx #1\expandafter \@firstoftwo
 \else \expandafter \@secondoftwo
 \fi
}%
\providecommand \natexlab [1]{#1}%
\providecommand \enquote  [1]{``#1''}%
\providecommand \bibnamefont  [1]{#1}%
\providecommand \bibfnamefont [1]{#1}%
\providecommand \citenamefont [1]{#1}%
\providecommand \href@noop [0]{\@secondoftwo}%
\providecommand \href [0]{\begingroup \@sanitize@url \@href}%
\providecommand \@href[1]{\@@startlink{#1}\@@href}%
\providecommand \@@href[1]{\endgroup#1\@@endlink}%
\providecommand \@sanitize@url [0]{\catcode `\\12\catcode `\$12\catcode
  `\&12\catcode `\#12\catcode `\^12\catcode `\_12\catcode `\%12\relax}%
\providecommand \@@startlink[1]{}%
\providecommand \@@endlink[0]{}%
\providecommand \url  [0]{\begingroup\@sanitize@url \@url }%
\providecommand \@url [1]{\endgroup\@href {#1}{\urlprefix }}%
\providecommand \urlprefix  [0]{URL }%
\providecommand \Eprint [0]{\href }%
\providecommand \doibase [0]{https://doi.org/}%
\providecommand \selectlanguage [0]{\@gobble}%
\providecommand \bibinfo  [0]{\@secondoftwo}%
\providecommand \bibfield  [0]{\@secondoftwo}%
\providecommand \translation [1]{[#1]}%
\providecommand \BibitemOpen [0]{}%
\providecommand \bibitemStop [0]{}%
\providecommand \bibitemNoStop [0]{.\EOS\space}%
\providecommand \EOS [0]{\spacefactor3000\relax}%
\providecommand \BibitemShut  [1]{\csname bibitem#1\endcsname}%
\let\auto@bib@innerbib\@empty
\bibitem [{\citenamefont {Tesch}\ and\ \citenamefont
  {de~Vivie-Riedle}(2002)}]{tesch2002quantum}%
  \BibitemOpen
  \bibfield  {author} {\bibinfo {author} {\bibfnamefont {C.~M.}\ \bibnamefont
  {Tesch}}\ and\ \bibinfo {author} {\bibfnamefont {R.}~\bibnamefont
  {de~Vivie-Riedle}},\ }\href
  {https://journals.aps.org/prl/abstract/10.1103/PhysRevLett.89.157901}
  {\bibfield  {journal} {\bibinfo  {journal} {Physical Review Letters}\
  }\textbf {\bibinfo {volume} {89}},\ \bibinfo {pages} {157901} (\bibinfo
  {year} {2002})}\BibitemShut {NoStop}%
\bibitem [{\citenamefont {Ashcroft}\ \emph {et~al.}(1976)\citenamefont
  {Ashcroft}, \citenamefont {Mermin} \emph {et~al.}}]{ashcroft1976solid}%
  \BibitemOpen
  \bibfield  {author} {\bibinfo {author} {\bibfnamefont {N.~W.}\ \bibnamefont
  {Ashcroft}}, \bibinfo {author} {\bibfnamefont {N.~D.}\ \bibnamefont
  {Mermin}}, \emph {et~al.},\ }\href@noop {} {\emph {\bibinfo {title} {Solid
  State Physics}}},\ Vol.\ \bibinfo {volume} {2005}\ (\bibinfo  {publisher}
  {Holt, Rinehart and Winston, New York London},\ \bibinfo {year}
  {1976})\BibitemShut {NoStop}%
\bibitem [{\citenamefont {Blais}\ \emph {et~al.}(2020)\citenamefont {Blais},
  \citenamefont {Girvin},\ and\ \citenamefont {Oliver}}]{blais2020quantum}%
  \BibitemOpen
  \bibfield  {author} {\bibinfo {author} {\bibfnamefont {A.}~\bibnamefont
  {Blais}}, \bibinfo {author} {\bibfnamefont {S.~M.}\ \bibnamefont {Girvin}},\
  and\ \bibinfo {author} {\bibfnamefont {W.~D.}\ \bibnamefont {Oliver}},\
  }\href {https://www.nature.com/articles/s41567-020-0806-z} {\bibfield
  {journal} {\bibinfo  {journal} {Nature Physics}\ }\textbf {\bibinfo {volume}
  {16}},\ \bibinfo {pages} {247} (\bibinfo {year} {2020})}\BibitemShut
  {NoStop}%
\bibitem [{\citenamefont {Chuang}\ \emph {et~al.}(1997)\citenamefont {Chuang},
  \citenamefont {Leung},\ and\ \citenamefont {Yamamoto}}]{chuang1997bosonic}%
  \BibitemOpen
  \bibfield  {author} {\bibinfo {author} {\bibfnamefont {I.~L.}\ \bibnamefont
  {Chuang}}, \bibinfo {author} {\bibfnamefont {D.~W.}\ \bibnamefont {Leung}},\
  and\ \bibinfo {author} {\bibfnamefont {Y.}~\bibnamefont {Yamamoto}},\ }\href
  {https://journals.aps.org/pra/abstract/10.1103/PhysRevA.56.1114} {\bibfield
  {journal} {\bibinfo  {journal} {Physical Review A}\ }\textbf {\bibinfo
  {volume} {56}},\ \bibinfo {pages} {1114} (\bibinfo {year}
  {1997})}\BibitemShut {NoStop}%
\bibitem [{\citenamefont {Gottesman}\ \emph {et~al.}(2001)\citenamefont
  {Gottesman}, \citenamefont {Kitaev},\ and\ \citenamefont
  {Preskill}}]{gottesman2001encoding}%
  \BibitemOpen
  \bibfield  {author} {\bibinfo {author} {\bibfnamefont {D.}~\bibnamefont
  {Gottesman}}, \bibinfo {author} {\bibfnamefont {A.}~\bibnamefont {Kitaev}},\
  and\ \bibinfo {author} {\bibfnamefont {J.}~\bibnamefont {Preskill}},\ }\href
  {https://journals.aps.org/pra/abstract/10.1103/PhysRevA.64.012310} {\bibfield
   {journal} {\bibinfo  {journal} {Physical Review A}\ }\textbf {\bibinfo
  {volume} {64}},\ \bibinfo {pages} {012310} (\bibinfo {year}
  {2001})}\BibitemShut {NoStop}%
\bibitem [{\citenamefont {Cochrane}\ \emph {et~al.}(1999)\citenamefont
  {Cochrane}, \citenamefont {Milburn},\ and\ \citenamefont
  {Munro}}]{cochrane1999macroscopically}%
  \BibitemOpen
  \bibfield  {author} {\bibinfo {author} {\bibfnamefont {P.~T.}\ \bibnamefont
  {Cochrane}}, \bibinfo {author} {\bibfnamefont {G.~J.}\ \bibnamefont
  {Milburn}},\ and\ \bibinfo {author} {\bibfnamefont {W.~J.}\ \bibnamefont
  {Munro}},\ }\href
  {https://journals.aps.org/pra/abstract/10.1103/PhysRevA.59.2631} {\bibfield
  {journal} {\bibinfo  {journal} {Physical Review A}\ }\textbf {\bibinfo
  {volume} {59}},\ \bibinfo {pages} {2631} (\bibinfo {year}
  {1999})}\BibitemShut {NoStop}%
\bibitem [{\citenamefont {Michael}\ \emph {et~al.}(2016)\citenamefont
  {Michael}, \citenamefont {Silveri}, \citenamefont {Brierley}, \citenamefont
  {Albert}, \citenamefont {Salmilehto}, \citenamefont {Jiang},\ and\
  \citenamefont {Girvin}}]{michael2016new}%
  \BibitemOpen
  \bibfield  {author} {\bibinfo {author} {\bibfnamefont {M.~H.}\ \bibnamefont
  {Michael}}, \bibinfo {author} {\bibfnamefont {M.}~\bibnamefont {Silveri}},
  \bibinfo {author} {\bibfnamefont {R.}~\bibnamefont {Brierley}}, \bibinfo
  {author} {\bibfnamefont {V.~V.}\ \bibnamefont {Albert}}, \bibinfo {author}
  {\bibfnamefont {J.}~\bibnamefont {Salmilehto}}, \bibinfo {author}
  {\bibfnamefont {L.}~\bibnamefont {Jiang}},\ and\ \bibinfo {author}
  {\bibfnamefont {S.~M.}\ \bibnamefont {Girvin}},\ }\href
  {https://journals.aps.org/prx/abstract/10.1103/PhysRevX.6.031006} {\bibfield
  {journal} {\bibinfo  {journal} {Physical Review X}\ }\textbf {\bibinfo
  {volume} {6}},\ \bibinfo {pages} {031006} (\bibinfo {year}
  {2016})}\BibitemShut {NoStop}%
\bibitem [{\citenamefont {Niu}\ \emph {et~al.}(2018{\natexlab{a}})\citenamefont
  {Niu}, \citenamefont {Chuang},\ and\ \citenamefont
  {Shapiro}}]{niu2018hardware}%
  \BibitemOpen
  \bibfield  {author} {\bibinfo {author} {\bibfnamefont {M.~Y.}\ \bibnamefont
  {Niu}}, \bibinfo {author} {\bibfnamefont {I.~L.}\ \bibnamefont {Chuang}},\
  and\ \bibinfo {author} {\bibfnamefont {J.~H.}\ \bibnamefont {Shapiro}},\
  }\href {https://journals.aps.org/pra/abstract/10.1103/PhysRevA.97.032323}
  {\bibfield  {journal} {\bibinfo  {journal} {Physical Review A}\ }\textbf
  {\bibinfo {volume} {97}},\ \bibinfo {pages} {032323} (\bibinfo {year}
  {2018}{\natexlab{a}})}\BibitemShut {NoStop}%
\bibitem [{\citenamefont {Albert}\ \emph {et~al.}(2018)\citenamefont {Albert},
  \citenamefont {Noh}, \citenamefont {Duivenvoorden}, \citenamefont {Young},
  \citenamefont {Brierley}, \citenamefont {Reinhold}, \citenamefont {Vuillot},
  \citenamefont {Li}, \citenamefont {Shen}, \citenamefont {Girvin} \emph
  {et~al.}}]{albert2018performance}%
  \BibitemOpen
  \bibfield  {author} {\bibinfo {author} {\bibfnamefont {V.~V.}\ \bibnamefont
  {Albert}}, \bibinfo {author} {\bibfnamefont {K.}~\bibnamefont {Noh}},
  \bibinfo {author} {\bibfnamefont {K.}~\bibnamefont {Duivenvoorden}}, \bibinfo
  {author} {\bibfnamefont {D.~J.}\ \bibnamefont {Young}}, \bibinfo {author}
  {\bibfnamefont {R.}~\bibnamefont {Brierley}}, \bibinfo {author}
  {\bibfnamefont {P.}~\bibnamefont {Reinhold}}, \bibinfo {author}
  {\bibfnamefont {C.}~\bibnamefont {Vuillot}}, \bibinfo {author} {\bibfnamefont
  {L.}~\bibnamefont {Li}}, \bibinfo {author} {\bibfnamefont {C.}~\bibnamefont
  {Shen}}, \bibinfo {author} {\bibfnamefont {S.}~\bibnamefont {Girvin}}, \emph
  {et~al.},\ }\href
  {https://journals.aps.org/pra/abstract/10.1103/PhysRevA.97.032346} {\bibfield
   {journal} {\bibinfo  {journal} {Physical Review A}\ }\textbf {\bibinfo
  {volume} {97}},\ \bibinfo {pages} {032346} (\bibinfo {year}
  {2018})}\BibitemShut {NoStop}%
\bibitem [{\citenamefont {Noh}\ \emph {et~al.}(2020)\citenamefont {Noh},
  \citenamefont {Girvin},\ and\ \citenamefont {Jiang}}]{noh2020encoding}%
  \BibitemOpen
  \bibfield  {author} {\bibinfo {author} {\bibfnamefont {K.}~\bibnamefont
  {Noh}}, \bibinfo {author} {\bibfnamefont {S.}~\bibnamefont {Girvin}},\ and\
  \bibinfo {author} {\bibfnamefont {L.}~\bibnamefont {Jiang}},\ }\href
  {https://journals.aps.org/prl/abstract/10.1103/PhysRevLett.125.080503}
  {\bibfield  {journal} {\bibinfo  {journal} {Physical Review Letters}\
  }\textbf {\bibinfo {volume} {125}},\ \bibinfo {pages} {080503} (\bibinfo
  {year} {2020})}\BibitemShut {NoStop}%
\bibitem [{\citenamefont {Gottesman}(1999)}]{gottesman1999fault}%
  \BibitemOpen
  \bibfield  {author} {\bibinfo {author} {\bibfnamefont {D.}~\bibnamefont
  {Gottesman}},\ }\href
  {https://www.sciencedirect.com/science/article/abs/pii/S0960077998002185}
  {\bibfield  {journal} {\bibinfo  {journal} {Chaos Solitons and Fractals}\
  }\textbf {\bibinfo {volume} {10}},\ \bibinfo {pages} {1749} (\bibinfo {year}
  {1999})}\BibitemShut {NoStop}%
\bibitem [{\citenamefont {Zhou}\ \emph {et~al.}(2003)\citenamefont {Zhou},
  \citenamefont {Zeng}, \citenamefont {Xu},\ and\ \citenamefont
  {Sun}}]{zhou2003quantum}%
  \BibitemOpen
  \bibfield  {author} {\bibinfo {author} {\bibfnamefont {D.}~\bibnamefont
  {Zhou}}, \bibinfo {author} {\bibfnamefont {B.}~\bibnamefont {Zeng}}, \bibinfo
  {author} {\bibfnamefont {Z.}~\bibnamefont {Xu}},\ and\ \bibinfo {author}
  {\bibfnamefont {C.}~\bibnamefont {Sun}},\ }\href
  {https://journals.aps.org/pra/abstract/10.1103/PhysRevA.68.062303} {\bibfield
   {journal} {\bibinfo  {journal} {Physical Review A}\ }\textbf {\bibinfo
  {volume} {68}},\ \bibinfo {pages} {062303} (\bibinfo {year}
  {2003})}\BibitemShut {NoStop}%
\bibitem [{\citenamefont {Niu}\ \emph {et~al.}(2018{\natexlab{b}})\citenamefont
  {Niu}, \citenamefont {Chuang},\ and\ \citenamefont {Shapiro}}]{niu2018qudit}%
  \BibitemOpen
  \bibfield  {author} {\bibinfo {author} {\bibfnamefont {M.~Y.}\ \bibnamefont
  {Niu}}, \bibinfo {author} {\bibfnamefont {I.~L.}\ \bibnamefont {Chuang}},\
  and\ \bibinfo {author} {\bibfnamefont {J.~H.}\ \bibnamefont {Shapiro}},\
  }\href {https://journals.aps.org/prl/abstract/10.1103/PhysRevLett.120.160502}
  {\bibfield  {journal} {\bibinfo  {journal} {Physical Review Letters}\
  }\textbf {\bibinfo {volume} {120}},\ \bibinfo {pages} {160502} (\bibinfo
  {year} {2018}{\natexlab{b}})}\BibitemShut {NoStop}%
\bibitem [{\citenamefont {Nielsen}\ and\ \citenamefont
  {Chuang}(2010)}]{nielsen2010quantum}%
  \BibitemOpen
  \bibfield  {author} {\bibinfo {author} {\bibfnamefont {M.~A.}\ \bibnamefont
  {Nielsen}}\ and\ \bibinfo {author} {\bibfnamefont {I.~L.}\ \bibnamefont
  {Chuang}},\ }\href
  {https://www.cambridge.org/core/books/quantum-computation-and-quantum-information/01E10196D0A682A6AEFFEA52D53BE9AE}
  {\emph {\bibinfo {title} {Quantum Computation and Quantum Information}}}\
  (\bibinfo  {publisher} {Cambridge University Press},\ \bibinfo {year}
  {2010})\BibitemShut {NoStop}%
\bibitem [{\citenamefont {Braunstein}\ and\ \citenamefont
  {Van~Loock}(2005)}]{braunstein2005quantum}%
  \BibitemOpen
  \bibfield  {author} {\bibinfo {author} {\bibfnamefont {S.~L.}\ \bibnamefont
  {Braunstein}}\ and\ \bibinfo {author} {\bibfnamefont {P.}~\bibnamefont
  {Van~Loock}},\ }\href
  {https://journals.aps.org/rmp/abstract/10.1103/RevModPhys.77.513} {\bibfield
  {journal} {\bibinfo  {journal} {Reviews of Modern Physics}\ }\textbf
  {\bibinfo {volume} {77}},\ \bibinfo {pages} {513} (\bibinfo {year}
  {2005})}\BibitemShut {NoStop}%
\bibitem [{\citenamefont {Pirandola}\ \emph {et~al.}(2006)\citenamefont
  {Pirandola}, \citenamefont {Mancini}, \citenamefont {Vitali},\ and\
  \citenamefont {Tombesi}}]{pirandola2006continuous}%
  \BibitemOpen
  \bibfield  {author} {\bibinfo {author} {\bibfnamefont {S.}~\bibnamefont
  {Pirandola}}, \bibinfo {author} {\bibfnamefont {S.}~\bibnamefont {Mancini}},
  \bibinfo {author} {\bibfnamefont {D.}~\bibnamefont {Vitali}},\ and\ \bibinfo
  {author} {\bibfnamefont {P.}~\bibnamefont {Tombesi}},\ }\href
  {https://link.springer.com/article/10.1140/epjd/e2005-00306-3} {\bibfield
  {journal} {\bibinfo  {journal} {The European Physical Journal D-Atomic,
  Molecular, Optical and Plasma Physics}\ }\textbf {\bibinfo {volume} {37}},\
  \bibinfo {pages} {283} (\bibinfo {year} {2006})}\BibitemShut {NoStop}%
\bibitem [{\citenamefont {Motes}\ \emph {et~al.}(2017)\citenamefont {Motes},
  \citenamefont {Baragiola}, \citenamefont {Gilchrist},\ and\ \citenamefont
  {Menicucci}}]{motes2017encoding}%
  \BibitemOpen
  \bibfield  {author} {\bibinfo {author} {\bibfnamefont {K.~R.}\ \bibnamefont
  {Motes}}, \bibinfo {author} {\bibfnamefont {B.~Q.}\ \bibnamefont
  {Baragiola}}, \bibinfo {author} {\bibfnamefont {A.}~\bibnamefont
  {Gilchrist}},\ and\ \bibinfo {author} {\bibfnamefont {N.~C.}\ \bibnamefont
  {Menicucci}},\ }\href
  {https://journals.aps.org/pra/abstract/10.1103/PhysRevA.95.053819} {\bibfield
   {journal} {\bibinfo  {journal} {Physical Review A}\ }\textbf {\bibinfo
  {volume} {95}},\ \bibinfo {pages} {053819} (\bibinfo {year}
  {2017})}\BibitemShut {NoStop}%
\bibitem [{\citenamefont {Fl{\"u}hmann}\ \emph {et~al.}(2019)\citenamefont
  {Fl{\"u}hmann}, \citenamefont {Nguyen}, \citenamefont {Marinelli},
  \citenamefont {Negnevitsky}, \citenamefont {Mehta},\ and\ \citenamefont
  {Home}}]{fluhmann2019encoding}%
  \BibitemOpen
  \bibfield  {author} {\bibinfo {author} {\bibfnamefont {C.}~\bibnamefont
  {Fl{\"u}hmann}}, \bibinfo {author} {\bibfnamefont {T.~L.}\ \bibnamefont
  {Nguyen}}, \bibinfo {author} {\bibfnamefont {M.}~\bibnamefont {Marinelli}},
  \bibinfo {author} {\bibfnamefont {V.}~\bibnamefont {Negnevitsky}}, \bibinfo
  {author} {\bibfnamefont {K.}~\bibnamefont {Mehta}},\ and\ \bibinfo {author}
  {\bibfnamefont {J.}~\bibnamefont {Home}},\ }\href
  {https://www.nature.com/articles/s41586-019-0960-6} {\bibfield  {journal}
  {\bibinfo  {journal} {Nature}\ }\textbf {\bibinfo {volume} {566}},\ \bibinfo
  {pages} {513} (\bibinfo {year} {2019})}\BibitemShut {NoStop}%
\bibitem [{\citenamefont {Ofek}\ \emph {et~al.}(2016)\citenamefont {Ofek},
  \citenamefont {Petrenko}, \citenamefont {Heeres}, \citenamefont {Reinhold},
  \citenamefont {Leghtas}, \citenamefont {Vlastakis}, \citenamefont {Liu},
  \citenamefont {Frunzio}, \citenamefont {Girvin}, \citenamefont {Jiang} \emph
  {et~al.}}]{ofek2016extending}%
  \BibitemOpen
  \bibfield  {author} {\bibinfo {author} {\bibfnamefont {N.}~\bibnamefont
  {Ofek}}, \bibinfo {author} {\bibfnamefont {A.}~\bibnamefont {Petrenko}},
  \bibinfo {author} {\bibfnamefont {R.}~\bibnamefont {Heeres}}, \bibinfo
  {author} {\bibfnamefont {P.}~\bibnamefont {Reinhold}}, \bibinfo {author}
  {\bibfnamefont {Z.}~\bibnamefont {Leghtas}}, \bibinfo {author} {\bibfnamefont
  {B.}~\bibnamefont {Vlastakis}}, \bibinfo {author} {\bibfnamefont
  {Y.}~\bibnamefont {Liu}}, \bibinfo {author} {\bibfnamefont {L.}~\bibnamefont
  {Frunzio}}, \bibinfo {author} {\bibfnamefont {S.}~\bibnamefont {Girvin}},
  \bibinfo {author} {\bibfnamefont {L.}~\bibnamefont {Jiang}}, \emph {et~al.},\
  }\href {https://www.nature.com/articles/nature18949} {\bibfield  {journal}
  {\bibinfo  {journal} {Nature}\ }\textbf {\bibinfo {volume} {536}},\ \bibinfo
  {pages} {441} (\bibinfo {year} {2016})}\BibitemShut {NoStop}%
\bibitem [{\citenamefont {Hastrup}\ \emph {et~al.}(2021)\citenamefont
  {Hastrup}, \citenamefont {Park}, \citenamefont {Brask}, \citenamefont
  {Filip},\ and\ \citenamefont {Andersen}}]{hastrup2021universal}%
  \BibitemOpen
  \bibfield  {author} {\bibinfo {author} {\bibfnamefont {J.}~\bibnamefont
  {Hastrup}}, \bibinfo {author} {\bibfnamefont {K.}~\bibnamefont {Park}},
  \bibinfo {author} {\bibfnamefont {J.~B.}\ \bibnamefont {Brask}}, \bibinfo
  {author} {\bibfnamefont {R.}~\bibnamefont {Filip}},\ and\ \bibinfo {author}
  {\bibfnamefont {U.~L.}\ \bibnamefont {Andersen}},\ }\href
  {https://arxiv.org/abs/2106.12272} {\bibfield  {journal} {\bibinfo  {journal}
  {arXiv preprint arXiv:2106.12272}\ } (\bibinfo {year} {2021})}\BibitemShut
  {NoStop}%
\bibitem [{\citenamefont {Shore}\ and\ \citenamefont
  {Knight}(1993)}]{shore1993jaynes}%
  \BibitemOpen
  \bibfield  {author} {\bibinfo {author} {\bibfnamefont {B.~W.}\ \bibnamefont
  {Shore}}\ and\ \bibinfo {author} {\bibfnamefont {P.~L.}\ \bibnamefont
  {Knight}},\ }\href
  {https://www.tandfonline.com/doi/abs/10.1080/09500349314551321} {\bibfield
  {journal} {\bibinfo  {journal} {Journal of Modern Optics}\ }\textbf {\bibinfo
  {volume} {40}},\ \bibinfo {pages} {1195} (\bibinfo {year}
  {1993})}\BibitemShut {NoStop}%
\bibitem [{\citenamefont {Law}\ and\ \citenamefont
  {Eberly}(1996)}]{law1996arbitrary}%
  \BibitemOpen
  \bibfield  {author} {\bibinfo {author} {\bibfnamefont {C.~K.}\ \bibnamefont
  {Law}}\ and\ \bibinfo {author} {\bibfnamefont {J.~H.}\ \bibnamefont
  {Eberly}},\ }\href
  {https://journals.aps.org/prl/abstract/10.1103/PhysRevLett.76.1055}
  {\bibfield  {journal} {\bibinfo  {journal} {Physical Review Letters}\
  }\textbf {\bibinfo {volume} {76}},\ \bibinfo {pages} {1055} (\bibinfo {year}
  {1996})}\BibitemShut {NoStop}%
\bibitem [{\citenamefont {Mischuck}\ and\ \citenamefont
  {M{\o}lmer}(2013)}]{mischuck2013qudit}%
  \BibitemOpen
  \bibfield  {author} {\bibinfo {author} {\bibfnamefont {B.}~\bibnamefont
  {Mischuck}}\ and\ \bibinfo {author} {\bibfnamefont {K.}~\bibnamefont
  {M{\o}lmer}},\ }\href
  {https://journals.aps.org/pra/abstract/10.1103/PhysRevA.87.022341} {\bibfield
   {journal} {\bibinfo  {journal} {Physical Review A}\ }\textbf {\bibinfo
  {volume} {87}},\ \bibinfo {pages} {022341} (\bibinfo {year}
  {2013})}\BibitemShut {NoStop}%
\bibitem [{\citenamefont {Strauch}(2012)}]{strauch2012all}%
  \BibitemOpen
  \bibfield  {author} {\bibinfo {author} {\bibfnamefont {F.~W.}\ \bibnamefont
  {Strauch}},\ }\href
  {https://journals.aps.org/prl/abstract/10.1103/PhysRevLett.109.210501}
  {\bibfield  {journal} {\bibinfo  {journal} {Physical Review Letters}\
  }\textbf {\bibinfo {volume} {109}},\ \bibinfo {pages} {210501} (\bibinfo
  {year} {2012})}\BibitemShut {NoStop}%
\bibitem [{\citenamefont {Santos}(2005)}]{santos2005universal}%
  \BibitemOpen
  \bibfield  {author} {\bibinfo {author} {\bibfnamefont {M.~F.}\ \bibnamefont
  {Santos}},\ }\href
  {https://journals.aps.org/prl/abstract/10.1103/PhysRevLett.95.010504}
  {\bibfield  {journal} {\bibinfo  {journal} {Physical Review Letters}\
  }\textbf {\bibinfo {volume} {95}},\ \bibinfo {pages} {010504} (\bibinfo
  {year} {2005})}\BibitemShut {NoStop}%
\bibitem [{\citenamefont {Krastanov}\ \emph {et~al.}(2015)\citenamefont
  {Krastanov}, \citenamefont {Albert}, \citenamefont {Shen}, \citenamefont
  {Zou}, \citenamefont {Heeres}, \citenamefont {Vlastakis}, \citenamefont
  {Schoelkopf},\ and\ \citenamefont {Jiang}}]{krastanov2015universal}%
  \BibitemOpen
  \bibfield  {author} {\bibinfo {author} {\bibfnamefont {S.}~\bibnamefont
  {Krastanov}}, \bibinfo {author} {\bibfnamefont {V.~V.}\ \bibnamefont
  {Albert}}, \bibinfo {author} {\bibfnamefont {C.}~\bibnamefont {Shen}},
  \bibinfo {author} {\bibfnamefont {C.-L.}\ \bibnamefont {Zou}}, \bibinfo
  {author} {\bibfnamefont {R.~W.}\ \bibnamefont {Heeres}}, \bibinfo {author}
  {\bibfnamefont {B.}~\bibnamefont {Vlastakis}}, \bibinfo {author}
  {\bibfnamefont {R.~J.}\ \bibnamefont {Schoelkopf}},\ and\ \bibinfo {author}
  {\bibfnamefont {L.}~\bibnamefont {Jiang}},\ }\href
  {https://journals.aps.org/pra/abstract/10.1103/PhysRevA.92.040303} {\bibfield
   {journal} {\bibinfo  {journal} {Physical Review A}\ }\textbf {\bibinfo
  {volume} {92}},\ \bibinfo {pages} {040303} (\bibinfo {year}
  {2015})}\BibitemShut {NoStop}%
\bibitem [{\citenamefont {Childs}\ and\ \citenamefont
  {Chuang}(2000)}]{childs2000universal}%
  \BibitemOpen
  \bibfield  {author} {\bibinfo {author} {\bibfnamefont {A.~M.}\ \bibnamefont
  {Childs}}\ and\ \bibinfo {author} {\bibfnamefont {I.~L.}\ \bibnamefont
  {Chuang}},\ }\href
  {https://journals.aps.org/pra/abstract/10.1103/PhysRevA.63.012306} {\bibfield
   {journal} {\bibinfo  {journal} {Physical Review A}\ }\textbf {\bibinfo
  {volume} {63}},\ \bibinfo {pages} {012306} (\bibinfo {year}
  {2000})}\BibitemShut {NoStop}%
\bibitem [{\citenamefont {Gulde}\ \emph {et~al.}(2003)\citenamefont {Gulde},
  \citenamefont {Riebe}, \citenamefont {Lancaster}, \citenamefont {Becher},
  \citenamefont {Eschner}, \citenamefont {H{\"a}ffner}, \citenamefont
  {Schmidt-Kaler}, \citenamefont {Chuang},\ and\ \citenamefont
  {Blatt}}]{gulde2003implementation}%
  \BibitemOpen
  \bibfield  {author} {\bibinfo {author} {\bibfnamefont {S.}~\bibnamefont
  {Gulde}}, \bibinfo {author} {\bibfnamefont {M.}~\bibnamefont {Riebe}},
  \bibinfo {author} {\bibfnamefont {G.~P.}\ \bibnamefont {Lancaster}}, \bibinfo
  {author} {\bibfnamefont {C.}~\bibnamefont {Becher}}, \bibinfo {author}
  {\bibfnamefont {J.}~\bibnamefont {Eschner}}, \bibinfo {author} {\bibfnamefont
  {H.}~\bibnamefont {H{\"a}ffner}}, \bibinfo {author} {\bibfnamefont
  {F.}~\bibnamefont {Schmidt-Kaler}}, \bibinfo {author} {\bibfnamefont {I.~L.}\
  \bibnamefont {Chuang}},\ and\ \bibinfo {author} {\bibfnamefont
  {R.}~\bibnamefont {Blatt}},\ }\href
  {https://www.nature.com/articles/nature01336?free=2} {\bibfield  {journal}
  {\bibinfo  {journal} {Nature}\ }\textbf {\bibinfo {volume} {421}},\ \bibinfo
  {pages} {48} (\bibinfo {year} {2003})}\BibitemShut {NoStop}%
\bibitem [{\citenamefont {Schmidt-Kaler}\ \emph {et~al.}(2003)\citenamefont
  {Schmidt-Kaler}, \citenamefont {H{\"a}ffner}, \citenamefont {Gulde},
  \citenamefont {Riebe}, \citenamefont {Lancaster}, \citenamefont {Deuschle},
  \citenamefont {Becher}, \citenamefont {H{\"a}nsel}, \citenamefont {Eschner},
  \citenamefont {Roos} \emph {et~al.}}]{schmidt2003realize}%
  \BibitemOpen
  \bibfield  {author} {\bibinfo {author} {\bibfnamefont {F.}~\bibnamefont
  {Schmidt-Kaler}}, \bibinfo {author} {\bibfnamefont {H.}~\bibnamefont
  {H{\"a}ffner}}, \bibinfo {author} {\bibfnamefont {S.}~\bibnamefont {Gulde}},
  \bibinfo {author} {\bibfnamefont {M.}~\bibnamefont {Riebe}}, \bibinfo
  {author} {\bibfnamefont {G.}~\bibnamefont {Lancaster}}, \bibinfo {author}
  {\bibfnamefont {T.}~\bibnamefont {Deuschle}}, \bibinfo {author}
  {\bibfnamefont {C.}~\bibnamefont {Becher}}, \bibinfo {author} {\bibfnamefont
  {W.}~\bibnamefont {H{\"a}nsel}}, \bibinfo {author} {\bibfnamefont
  {J.}~\bibnamefont {Eschner}}, \bibinfo {author} {\bibfnamefont
  {C.}~\bibnamefont {Roos}}, \emph {et~al.},\ }\href
  {https://link.springer.com/article/10.1007/s00340-003-1346-9} {\bibfield
  {journal} {\bibinfo  {journal} {Applied Physics B}\ }\textbf {\bibinfo
  {volume} {77}},\ \bibinfo {pages} {789} (\bibinfo {year} {2003})}\BibitemShut
  {NoStop}%
\bibitem [{\citenamefont {Vandersypen}\ and\ \citenamefont
  {Chuang}(2005)}]{vandersypen2005nmr}%
  \BibitemOpen
  \bibfield  {author} {\bibinfo {author} {\bibfnamefont {L.~M.}\ \bibnamefont
  {Vandersypen}}\ and\ \bibinfo {author} {\bibfnamefont {I.~L.}\ \bibnamefont
  {Chuang}},\ }\href
  {https://journals.aps.org/rmp/abstract/10.1103/RevModPhys.76.1037} {\bibfield
   {journal} {\bibinfo  {journal} {Reviews of Modern Physics}\ }\textbf
  {\bibinfo {volume} {76}},\ \bibinfo {pages} {1037} (\bibinfo {year}
  {2005})}\BibitemShut {NoStop}%
\bibitem [{\citenamefont {Pryor}\ and\ \citenamefont
  {Khaneja}(2006)}]{pryor2006fourier}%
  \BibitemOpen
  \bibfield  {author} {\bibinfo {author} {\bibfnamefont {B.}~\bibnamefont
  {Pryor}}\ and\ \bibinfo {author} {\bibfnamefont {N.}~\bibnamefont
  {Khaneja}},\ }\href {https://aip.scitation.org/doi/abs/10.1063/1.2390715}
  {\bibfield  {journal} {\bibinfo  {journal} {The Journal of Chemical Physics}\
  }\textbf {\bibinfo {volume} {125}},\ \bibinfo {pages} {194111} (\bibinfo
  {year} {2006})}\BibitemShut {NoStop}%
\bibitem [{\citenamefont {Goldberg}\ and\ \citenamefont
  {Steinberg}(2020)}]{goldberg2020transcoherent}%
  \BibitemOpen
  \bibfield  {author} {\bibinfo {author} {\bibfnamefont {A.~Z.}\ \bibnamefont
  {Goldberg}}\ and\ \bibinfo {author} {\bibfnamefont {A.~M.}\ \bibnamefont
  {Steinberg}},\ }\href {https://link.aps.org/doi/10.1103/PRXQuantum.1.020306}
  {\bibfield  {journal} {\bibinfo  {journal} {PRX Quantum}\ }\textbf {\bibinfo
  {volume} {1}},\ \bibinfo {pages} {020306} (\bibinfo {year}
  {2020})}\BibitemShut {NoStop}%
\bibitem [{\citenamefont {Brennen}\ \emph {et~al.}(2005)\citenamefont
  {Brennen}, \citenamefont {O’Leary},\ and\ \citenamefont
  {Bullock}}]{brennen2005criteria}%
  \BibitemOpen
  \bibfield  {author} {\bibinfo {author} {\bibfnamefont {G.~K.}\ \bibnamefont
  {Brennen}}, \bibinfo {author} {\bibfnamefont {D.~P.}\ \bibnamefont
  {O’Leary}},\ and\ \bibinfo {author} {\bibfnamefont {S.~S.}\ \bibnamefont
  {Bullock}},\ }\href
  {https://journals.aps.org/pra/abstract/10.1103/PhysRevA.71.052318} {\bibfield
   {journal} {\bibinfo  {journal} {Physical Review A}\ }\textbf {\bibinfo
  {volume} {71}},\ \bibinfo {pages} {052318} (\bibinfo {year}
  {2005})}\BibitemShut {NoStop}%
\bibitem [{\citenamefont {Bertlmann}\ and\ \citenamefont
  {Krammer}(2008)}]{bertlmann2008bloch}%
  \BibitemOpen
  \bibfield  {author} {\bibinfo {author} {\bibfnamefont {R.~A.}\ \bibnamefont
  {Bertlmann}}\ and\ \bibinfo {author} {\bibfnamefont {P.}~\bibnamefont
  {Krammer}},\ }\href
  {https://iopscience.iop.org/article/10.1088/1751-8113/41/23/235303/meta}
  {\bibfield  {journal} {\bibinfo  {journal} {Journal of Physics A:
  Mathematical and Theoretical}\ }\textbf {\bibinfo {volume} {41}},\ \bibinfo
  {pages} {235303} (\bibinfo {year} {2008})}\BibitemShut {NoStop}%
\bibitem [{\citenamefont {Mintzer}(2021)}]{mintzer2021repo}%
  \BibitemOpen
  \bibfield  {author} {\bibinfo {author} {\bibfnamefont {G.}~\bibnamefont
  {Mintzer}},\ }\href@noop {} {\bibinfo {title} {qo-qudit}},\ \bibinfo
  {howpublished} {\url{https://github.com/mitquanta/qo-qudit}} (\bibinfo {year}
  {2021})\BibitemShut {NoStop}%
\bibitem [{\citenamefont {Lidar}\ \emph {et~al.}(1998)\citenamefont {Lidar},
  \citenamefont {Chuang},\ and\ \citenamefont {Whaley}}]{lidar1998decoherence}%
  \BibitemOpen
  \bibfield  {author} {\bibinfo {author} {\bibfnamefont {D.~A.}\ \bibnamefont
  {Lidar}}, \bibinfo {author} {\bibfnamefont {I.~L.}\ \bibnamefont {Chuang}},\
  and\ \bibinfo {author} {\bibfnamefont {K.~B.}\ \bibnamefont {Whaley}},\
  }\href {https://journals.aps.org/prl/abstract/10.1103/PhysRevLett.81.2594}
  {\bibfield  {journal} {\bibinfo  {journal} {Physical Review Letters}\
  }\textbf {\bibinfo {volume} {81}},\ \bibinfo {pages} {2594} (\bibinfo {year}
  {1998})}\BibitemShut {NoStop}%
\end{thebibliography}%
\end{document}